\documentclass[11pt, twocolumn]{aastex631}
\usepackage{multirow}
\usepackage{lineno}
\newcommand{\kms}{${\rm km~s^{-1}}$}
\newcommand{\Moyr}{$M_{\odot}~{\rm yr^{-1}}$}
\newcommand{\Ha}{${\rm H\alpha}$}
\newcommand{\Hb}{${\rm H\beta}$}

\begin{document}
	
\title{\textbf{\large{A GMOS/IFU Study of 
Jellyfish Galaxies in Massive Clusters 
}}}

\author[0000-0003-3301-759X]{Jeong Hwan Lee}
\affil{Astronomy Program, Department of Physics and Astronomy, SNUARC, Seoul National University, 1 Gwanak-ro, Gwanak-gu, Seoul 08826, Republic of Korea}
\author[0000-0003-2713-6744]{Myung Gyoon Lee}
\affil{Astronomy Program, Department of Physics and Astronomy, SNUARC, Seoul National University, 1 Gwanak-ro, Gwanak-gu, Seoul 08826, Republic of Korea}
\author[0000-0002-3706-9955]{Jae Yeon Mun}
\affil{Research School of Astronomy and Astrophysics, Australian National University, Canberra, ACT 2611, Australia}
\author{Brian S. Cho}
\affil{Astronomy Program, Department of Physics and Astronomy, SNUARC, Seoul National University, 1 Gwanak-ro, Gwanak-gu, Seoul 08826, Republic of Korea}
\author[0000-0003-3734-1995]{Jisu Kang}
\affil{Astronomy Program, Department of Physics and Astronomy, SNUARC, Seoul National University, 1 Gwanak-ro, Gwanak-gu, Seoul 08826, Republic of Korea}

\correspondingauthor{Myung Gyoon Lee}
\email{joungh93@snu.ac.kr, mglee@astro.snu.ac.kr}

\begin{abstract}
Jellyfish galaxies are an intriguing snapshot of galaxies undergoing ram-pressure stripping (RPS) in dense environments, showing spectacular star-forming knots in their disks and tails.
We study the ionized gas properties of five jellyfish galaxies in massive clusters with Gemini GMOS/IFU observations: MACSJ0916-JFG1 ($z=0.330$), MACSJ1752-JFG2 ($z=0.353$), A2744-F0083 ($z=0.303$), MACSJ1258-JFG1 ($z=0.342$), and MACSJ1720-JFG1 ($z=0.383$).
BPT diagrams show that various mechanisms (star formation, AGN, or mixed effects) are ionizing gas in these 
galaxies. 
Radial velocity distributions of ionized gas seem to 
follow disk rotation of galaxies, with the appearance of 
a few high-velocity components in the tails as a sign of RPS.
Mean gas velocity dispersion is lower than 50 \kms~in most star-forming regions except near AGNs or shock-heated regions, 
indicating that the ionized gas 
is dynamically cold. 
Integrated star formation rates (SFRs) of these 
galaxies range from $7~{\rm M_{\odot}~{\rm yr^{-1}}}$ to $35~{\rm M_{\odot}~{\rm yr^{-1}}}$ and the tail SFRs 
are from $0.6~{\rm M_{\odot}~{\rm yr^{-1}}}$ to $16~{\rm M_{\odot}~{\rm yr^{-1}}}$, which are much higher than those of other jellyfish galaxies in the local universe.
These high SFR values imply that RPS triggers intense star formation activity in these extreme jellyfish 
galaxies.
The phase-space diagrams demonstrate that the jellyfish galaxies with higher stellar masses and higher host cluster velocity dispersion are likely to have more enhanced star formation activity.
The jellyfish galaxies in this study have similar gas kinematics and dynamical states to those in the local universe, but they show a much higher SFR. 
\end{abstract}


\section{Introduction}


Environmental effects play an important role in galaxy transformation 
associated with high-density environments.
Most gas-rich galaxies in galaxy groups or clusters evolve by losing their gas content due to various mechanisms: galaxy mergers \citep{tom72}, ram-pressure stripping \citep[RPS;][]{gun72}, galaxy harassment \citep{moo96}, and starvation \citep{lar80}.
These external processes remove the gas ingredients for star formation from galaxies and transform late-type galaxies into early-type galaxies in dense environments.

Among these mechanisms, RPS is the interaction between the intracluster medium (ICM) and the interstellar medium (ISM) in galaxies.
It has been known as the most efficient mechanism of gas removal from galaxies in cluster environments.
When a galaxy moves through the intergalactic space that is filled with hot and diffuse gas, ram pressure applies a force to the gas component in the galaxy towards the direction opposite of the galaxy motion.
RPS eventually quenches star formation activity by sweeping out the gas content in galaxies.
Since diffuse ICM gas with X-ray emission is frequently observed in galaxy clusters, RPS seems to significantly contribute to the passive evolution of cluster galaxies \citep{dre80}.

However, some late-type galaxies undergoing RPS are not quiescent but rather show local star-forming regions.
Several late-type galaxies under the effect of RPS in Virgo and Coma were found to have a large number of blue and UV-emitting knots or ``spur clusters'' throughout their disks \citep{hes10, smi10, yos12, lee16}.
Interestingly, a post-merger elliptical galaxy in Abell 2670 was revealed to be undergoing RPS and exhibits compact star-forming blobs outside the main body of the galaxy \citep{she17}.
Cluster galaxies with remarkable RPS features like one-sided tails and bright star-forming knots 
are often called ``jellyfish galaxies'' \citep{bek09, chu09}.
The presence of young stellar systems in jellyfish galaxies implies that RPS can locally boost the star formation activity in gas-rich galaxies before removing gas completely.
Thus, jellyfish galaxies are very interesting and useful targets to investigate the relationship between star formation activity and RPS.

\begin{deluxetable*}{lccc}[h]
	\tabletypesize{\footnotesize}
	\setlength{\tabcolsep}{0.05in}
	\tablecaption{Examples of Jellyfish Galaxies (JFGs) 
	at $z>0.1$}
	\tablehead{\colhead{Reference} & \colhead{JFGs (Host System)} & \colhead{Redshift} & \colhead{Observing Instruments}}
	\startdata
	\citet{owe06} & C153 (A2125) & $z=0.253$ & \textit{HST}/WFPC2, KPNO 4m, \\
	& & & Gemini/GMOS longslit \\
	& & & VLA, \textit{Chandra} \\ \hline
	\citet{cor07} & 131124$-$012040 (A1689) & $z=0.187$ & \textit{HST}/WFPC2, \textit{HST}/ACS, \\
	& 235144$-$260358 (A2667) & $z=0.227$ & VLT/ISAAC, \textit{Spitzer}/IRAC, \\
	& & & \textit{Spitzer}/MIPS, VLA, \\
	& & & VLT/VIMOS, Keck/LRIS \\ \hline
	\citet{owe12} & 4 JFGs (A2744) & $z=0.29-0.31$ & \textit{HST}/ACS \\
	& & & AAT/AAOmega \\ \hline
	\citet{ebe14} & 6 JFGs & $z=0.3-0.5$ & \textit{HST}/ACS \\ 
	& (37 MACS clusters) & & \\ \hline
	\citet{mcp16} & 16 JFGs & $z=0.3-0.5$ & \textit{HST}/ACS, \\
	& (63 MACS clusters) & & Keck/DEIMOS \\ \hline
	\citet{ebe17} & MACSJ0553-JFG1 & $z=0.442$ & \textit{HST}/ACS, \textit{HST}/WFC3, \\
	& (MACSJ0553$-$3342) & & Keck/LRIS, Keck/DEIMOS, \\
	& & & \textit{Chandra}, GMRT \\ \hline
	\citet{bos19} & ID 345 and ID 473 (CGr32) & $z=0.73$ & \textbf{VLT/MUSE\tablenotemark{\rm \footnotesize a}}, \textit{XMM-Newton}, \\
	& & & \textit{HST}/ACS, Subaru/Suprime-Cam \\ \hline
	\citet{kal19} & A1758N\_JFG1 (A1758N) & $z=0.273$ & \textbf{Keck/KCWI\tablenotemark{\rm \footnotesize a}}, Keck/DEIMOS, \\
	& & & \textit{HST}/ACS, \textit{Chandra} \\ \hline
	\citet{rom19} & 70 JFGs (A901/2) & $z=0.165$ & GTC/OSIRIS (the OMEGA survey),  \\
	& & & \textit{HST}/ACS, \textit{XMM-Newton} \\ \hline
	\citet{cas20} & The southern companion of BCG & $z=1.71$ & IRAM/NOEMA, \textit{HST}/ACS \\
	& (SpARCS1049+56) & & \\ \hline
	\citet{dur21} & 81 JFGs (MACS0717.5$+$3745) & $z=0.546$ & \textit{HST}/ACS \\
	& 97 JFGs (22 clusters) & $z=0.2-0.9$ & NASA/IPAC Extragalactic Database \\ \hline
	\citet{mor22} & 13 JFGs (A2744 and A370) & $z=0.3-0.4$ & \textit{HST}/ACS, \textbf{VLT/MUSE\tablenotemark{\rm \footnotesize a}}
	\enddata
	\label{tab1}
	\tablenotetext{}{\textbf{Note.}}
	\tablenotetext{\rm a}{Integral field spectroscopic (IFS) observations.}
\end{deluxetable*}

Recently, observational studies on jellyfish galaxies have utilized integral field spectroscopy (IFS).
IFS observations 
provide both spatial and spectral information, so they are optimal to investigate various physical properties of jellyfish galaxies.
For instance, the GAs Stripping Phenomena (GASP) survey observed a large sample of jellyfish galaxies in nearby galaxy clusters using the Multi Unit Spectroscopic Explorer (MUSE) on the Very Large Telescope (VLT) \citep{pog17}.
Studies from the GASP survey found that star-forming knots in jellyfish galaxies are formed in situ with stellar ages younger than 10 Myr, 
are dynamically cold ($\sigma_{v,{\rm gas}}<40~{\rm km~s^{-1}}$), and are mostly ionized by photons emitted from young massive stars 
\citep{bel17, bel19}.
They obtained a mean star formation rate (SFR) of $\sim1.8$ \Moyr~in the disks and $\sim0.13$ \Moyr~in the tails of the jellyfish galaxies \citep{pog19, gul20}.
As a result, the GASP survey has successfully provided a comprehensive view of the star formation activity of jellyfish galaxies in the local universe ($z<0.1$).

However, most IFS studies have been limited to jellyfish galaxies in low-mass clusters ($\sigma_{v,{\rm cl}}\lesssim1000$ \kms) in the local universe ($z<0.1$).
Since massive clusters are rare in the nearby universe, jellyfish galaxies in massive clusters have been mostly found at intermediate redshift $(z>0.1)$ and were studied 
using deep images from the Hubble Space Telescope (\textit{HST}).
{\color{blue} {\bf Table \ref{tab1}}} lists the jellyfish galaxies that have been mainly observed at $z>0.1$ so far.
All the observations used \textit{HST} optical images because such high-resolution images show the dramatic features of jellyfish galaxies very well.
\citet{owe06} and \citet{cor07} found three disturbed jellyfish galaxies with blue star-forming knots in the multi-band \textit{HST} optical images and radio images: C153 (Abell 2125; $z=0.253$), 131124-012040 (Abell 1689; $z=0.187$), and 235144-260358 (Abell 2667; $z=0.227$).
\citet{owe12} studied four jellyfish galaxies with highly asymmetric tails and very bright star-forming knots in the merging cluster Abell 2744, using the \textit{HST} optical images and AAOmega spectra.
\citet{ebe14} and \citet{mcp16} performed a systematic search of jellyfish galaxies using the \textit{HST} images of the Massive Cluster Survey \citep[MACS;][]{ebe01,ebe10} and provided a catalog of 16 jellyfish galaxies in the MACS cluster samples at $z>0.3$ ($\langle M_{\rm tot}\rangle\sim1.3\times10^{15}~M_{\odot}$).
\citet{rom19} presented the characteristics of 70 jellyfish galaxies using the OMEGA-OSIRIS survey of a multi-cluster system A901/2.
Similarly, \citet{dur21} found a total of 178 jellyfish galaxy candidates in 23 clusters from the DAFT/FADA and CLASH surveys.

Among the studies described 
in {\color{blue} {\bf Table \ref{tab1}}}, the IFS observations of jellyfish galaxies at $z>0.1$ 
 are rare except for the three cases.
\citet{bos19} detected two massive star-forming galaxies with long gaseous tails in the COSMOS cluster CGr32 ($z=0.73$) using VLT/MUSE.
\citet{kal19} studied an extreme jellyfish galaxy, A1758N-JFG1, in the colliding galaxy cluster Abell 1758N ($z=0.28)$ with the Keck Cosmic Web Imager (KCWI).
These two studies focused on the resolved kinematics of jellyfish galaxies based on the [\ion{O}{2}]$\lambda\lambda3727,3729$ doublet emission line because their IFS data did not cover the \Ha~emission line.
Recently, \citet{mor22} presented the properties of 13 jellyfish galaxies in the inner regions of Abell 2744 ($z=0.306$) and Abell 370 ($z=0.375$) using the emission lines of \Ha~and [\ion{O}{2}]$\lambda\lambda3727,3729$.
Their study suggested that the [\ion{O}{2}]/\Ha~ratio in the tails of their sample galaxies is much higher than those of local jellyfish galaxies, which implies that the stripped gas might have low gas density or interact with the ICM.


In this study, we investigate the physical properties of jellyfish galaxies in massive clusters using the integral field unit (IFU) instrument on the Gemini Multi-Object Spectrograph (GMOS).
The 8-m Gemini telescope is suitable for observing jellyfish galaxies at intermediate redshift in terms of its light-gathering power and field of view (FOV).
We carry out emission line analyses of the GMOS/IFU data of jellyfish galaxies, which includes the \Ha~line.
%
Using the \Ha-derived SFRs of jellyfish galaxies in this study, 
\citet{lee22} presented a relation between the star formation activity of the jellyfish galaxies and the host cluster properties. 
In this paper, we present the detailed methods and analyses of emission lines of the jellyfish galaxies.

This paper is organized as follows.
In {\color{blue} {\bf Section \ref{sec:samp}}}, we explain our target selection and their properties.
Then, we describe our GMOS/IFU observations and data reduction in {\color{blue} {\bf Section \ref{sec:gemini}}}, and we describe the multi-wavelength archival images in {\color{blue} {\bf Section \ref{sec:image}}}.
We explain our analysis of the GMOS/IFU data cube to derive the physical quantities in {\color{blue} {\bf Section \ref{sec:analysis}}}.
In {\color{blue} {\bf Section \ref{sec:result}}}, we illustrate the maps of the gas ionization mechanisms, the kinematics, the star formation activity, and the dynamical states for our targets.
In {\color{blue} {\bf Section \ref{sec:psd}}}, we discuss the star formation activity of jellyfish galaxies in this study in comparison with those in the literature 
using phase-space diagrams.
Finally, we summarize 
primary results in {\color{blue} {\bf Section \ref{sec:summary}}}.
We adopt the cosmological parameters with $H_{0}=70~{\rm km~s^{-1}~Mpc^{-1}}$, $\Omega_{M}=0.3$, and $\Omega_{\Lambda}=0.7$ in this study.



\begin{figure*}[h]
	\centering
	\includegraphics[width=0.925\textwidth]{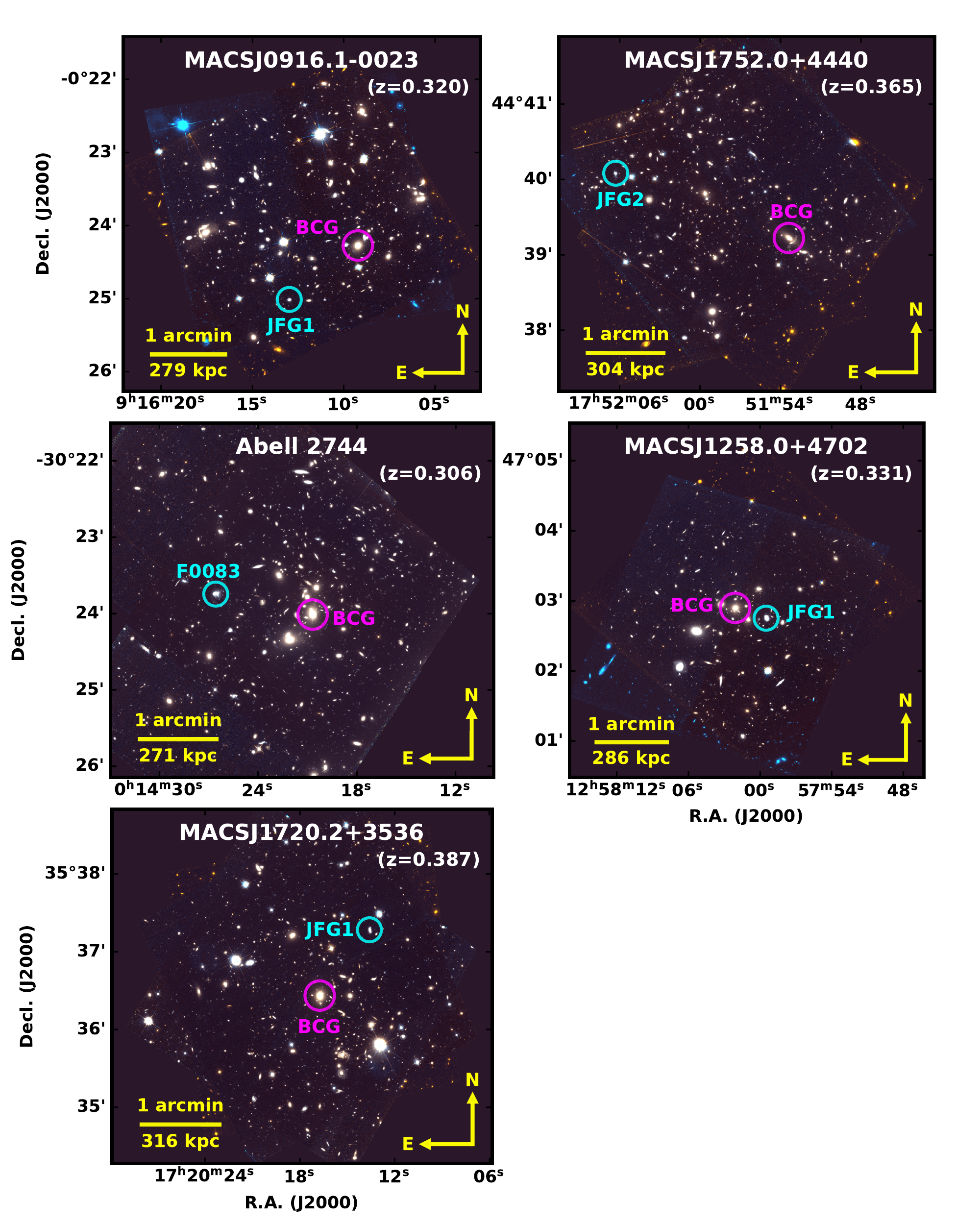}
	\caption{
	\textit{HST} color images of the host clusters of the jellyfish galaxies in this study.
    These images were prepared by 
    combining the two band images of \textit{HST}/ACS $F606W$ and $F814W$.
    Cyan circles denote the location of each jellyfish galaxy.
    Magenta circles denote the brightest cluster galaxy (BCG), which is set to be the center of each cluster.
    The orientations and the angular scales are marked.
	\label{fig:HST1}}
\end{figure*}

\section{Sample and Data}

\subsection{Jellyfish Galaxy Sample}
\label{sec:samp}

\begin{deluxetable*}{lccccc}
	\tabletypesize{\footnotesize}
	\setlength{\tabcolsep}{0.15in}
	\tablecaption{Properties of the Host Clusters of the Jellyfish Galaxies}
	\tablehead{\colhead{Cluster Name\tablenotemark{\rm \footnotesize a}} & \colhead{M0916} & \colhead{M1752} & \colhead{A2744} & \colhead{M1258} & \colhead{M1720}}
	\startdata
	R.A. (J2000)\tablenotemark{\rm \footnotesize b} & 09:16:13.9 & 17:52:01.5 & 00:14:18.9 & 12:58:02.1 & 17:20:16.8 \\
	Decl. (J2000)\tablenotemark{\rm \footnotesize b} & $-$00:24:42 & $+$44:40:46 & $-$30:23:22 & $+$47:02:54 & $+$35:36:26 \\
	Redshift ($z$)\tablenotemark{\rm \footnotesize c} & 0.320 & 0.365 & 0.306 & 0.331 & 0.387 \\
	Luminosity distance (Mpc)\tablenotemark{\rm \footnotesize d} & 1673.0 & 1949.1 & 1591.0 & 1740.0 & 2089.1 \\
	Angular scale (${\rm kpc~arcsec^{-1}}$)\tablenotemark{\rm \footnotesize d} & 4.655 & 5.073 & 4.519 & 4.762 & 5.265 \\
	Age at redshift (Gyr)\tablenotemark{\rm \footnotesize d} & 9.86 & 9.47 & 9.98 & 9.76 & 9.29 \\
	$E(B-V)$ (mag)\tablenotemark{\rm \footnotesize e} & 0.027 & 0.027 & 0.012 & 0.012 & 0.034 \\
	Velocity dispersion (\kms)\tablenotemark{\rm \footnotesize f} & $1066^{\ast}$ & 1186 & 1301 & $872^{\ast}$ & $1068^{\ast}$ \\
	Virial mass ($10^{14}~M_{\odot}$)\tablenotemark{\rm \footnotesize g} & 11.6 & $15.5^{\ast}$ & 22.0 & 6.3 & 11.2 \\
	Virial radius (Mpc)\tablenotemark{\rm \footnotesize g} & 1.9 & 2.1 & 2.4 & 1.6 & 1.9 \\
	X-ray luminosity (${\rm erg~s^{-1}}$)\tablenotemark{\rm \footnotesize h} & $5.9\times10^{44}$ & $6.3\times10^{44}$ & $1.6\times10^{45}$ & $3.5\times10^{44}$ & $1.6\times10^{45}$
	\enddata
	\label{tab2}
	\tablenotetext{}{\textbf{Notes.}}
	\tablenotetext{\rm a}{MACSJ0916.1$-$0023 (M0916), MACSJ1752.0$+$4440 (M1752), Abell 2744 (A2744), MACSJ1258.0$+$4702 (M1258), and MACSJ1720.2$+$3536 (M1720).}
	\tablenotetext{\rm b}{NASA/IPAC Extragalactic Database (NED).}
	\tablenotetext{\rm c}{\citet{rep18} for MACSJ0916.1$-$0023 and MACSJ1258.0$+$4702, \citet{gol19} for MACSJ1752$+$4440, \citet{owe11} for Abell 2744, and \citet{ebe10} for MACSJ1720.2$+$3536.}
	\tablenotetext{\rm d}{Based on the cosmological parameters of $H_{0}=70~{\rm km~s^{-1}~Mpc^{-1}}$, $\Omega_{M}=0.3$, $\Omega_{\rm \Lambda}=0.7$, and $T_{\rm CMB}=2.73~{\rm K}$.}
	\tablenotetext{\rm e}{Foreground reddening from \citet{sch11}.}
	\tablenotetext{\rm f}{\citet{gol19} for MACSJ1752$+$4440, \citet{bos06} for the main clump of Abell 2744, and $^{\ast}$ marks the velocity dispersion estimated from the scaling relation of \citet{evr08}.}
	\tablenotetext{\rm g}{
	Virial mass ($M_{200}$) and virial radius ($r_{200}$) are from the following literature: \citet{mede18} and \citet{osa20} for MACSJ0916$-$0023, \citet{gol19} for MACSJ1752.0$+$4440, and \citet{bos06} for Abell 2744, \citet{ser17} for MACSJ1258.0$+$4702, and \citet{ume18} for MACSJ1720.2$+$3536.
	Asterisks mark the virial mass estimated from the scaling relation of \citet{evr08}.}
	\tablenotetext{\rm h}{The ROSAT all-sky survey \citep[$0.1-2.4~{\rm keV}$;][]{vog99}.}
\end{deluxetable*}

\begin{figure*}
	\centering
	\includegraphics[width=\textwidth]{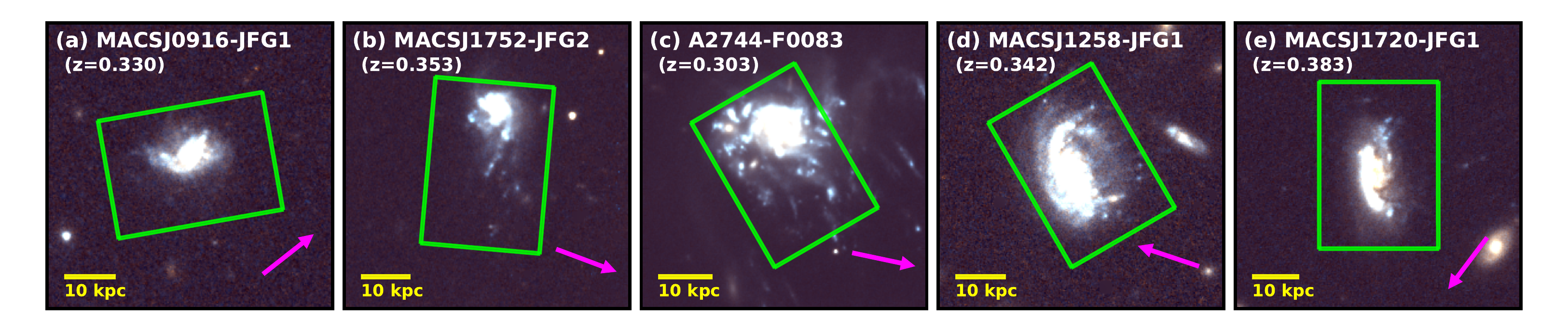}
	\caption{
	The zoom-in \textit{HST} images of the five jellyfish galaxies with the  size of $12''\times12''$.
	Green boxes show the field-of-view of the GMOS/IFU 2-slit mode ($5''\times7''$).
	Magenta arrows represent the directions to the cluster center (BCGs).
	The orientations are same as in {\color{blue} {\bf Figure \ref{fig:HST1}}}.
	\label{fig:HST2}}
\end{figure*}

We selected five jellyfish galaxies in the MACS clusters \citep{mcp16} and Abell 2744 \citep{owe12} for our IFS observations: MACSJ0916-JFG1 ($z=0.330$), MACSJ1752-JFG2 ($z=0.353$), A2744-F0083 ($z=0.303$), MACSJ1258-JFG1 ($z=0.342$), and MACSJ1720-JFG1 ($z=0.383$).
Among these five galaxies, A2744-F0083 and MACSJ1258-JFG1 host type I active galactic nuclei (AGN) in their center.
We will describe this again with the Baldwin, Phillips \& Terlevich (BPT) diagrams \citep{bal81} in {\color{blue} {\bf Section \ref{sec:bpt}}}.
The properties of the host clusters are summarized in {\color{blue} {\bf Table \ref{tab2}}}.

{\color{blue} {\bf Figure \ref{fig:HST1}}} shows the \textit{HST}/ACS color composite images of the selected jellyfish galaxies and their host clusters.
We display the locations of the brightest cluster galaxies (BCGs) and jellyfish galaxies in the host clusters.
In this study, we set the marked BCGs as the center of each cluster.
{\color{blue} {\bf Figure \ref{fig:HST2}}} shows the zoom-in thumbnail images of the five jellyfish galaxies.
We mark the FOV of the GMOS/IFU 2-slit mode (green boxes; $5\arcsec\times7\arcsec$) and the direction to the cluster center (magenta arrows).
In these \textit{HST} optical images, all the jellyfish galaxies show disturbed and asymmetric morphology with blue tails and bright star-forming knots.
MACSJ0916-JFG1 has a short ($\sim5~{\rm kpc}$) tail at the eastern side of the galactic center and a few star-forming knots in the disk and tail.
MACSJ1752-JFG2 shows not only an extended ($\sim25~{\rm kpc}$) tail towards the southern direction, but also large bright knots with ongoing star formation around the disk.
A2744-F0083 is the most spectacular object exhibiting large and bright star-forming knots and multiple stripped tails.
MACSJ1258-JFG1 has a bright tail towards the northern direction with a length of $\sim10~{\rm kpc}$ and small blue extraplanar knots outside the disk.
MACSJ1720-JFG1 is the most distant galaxy in this study.
This galaxy shows a star-forming tail in the northern region and a bright compressed region at its head.
The FOV of the GMOS/IFU is wide enough to cover the substructures of the jellyfish galaxies except for a few faint tails at the western side of A2744-F0083.

For these jellyfish galaxies, there seems to be no clear trend 
between the direction of the tails and their direction towards the cluster center.
Tails of MACSJ0916-JFG1 and MACSJ1752-JFG2 are extended in a direction nearly tangential to the cluster center.
In contrast, tails of MACSJ1258-JFG1 and MACSJ1720-JFG1 are extended towards the opposite direction to the cluster center.
Interestingly, A2744-F0083 has multiple tails extending towards the BCGs in the cluster center.
This misalignment of the direction of tails and the direction to the cluster center 
has been also shown by the jellyfish galaxies in the GASP survey \citep{pog16} and those in the IllustrisTNG simulation \citep{yun19}.
This indicates that these jellyfish galaxies travel in various orbital trajectories within their host clusters, and do not necessarily travel in a radial direction.


\begin{figure*}
	\centering
	\includegraphics[width=\textwidth]{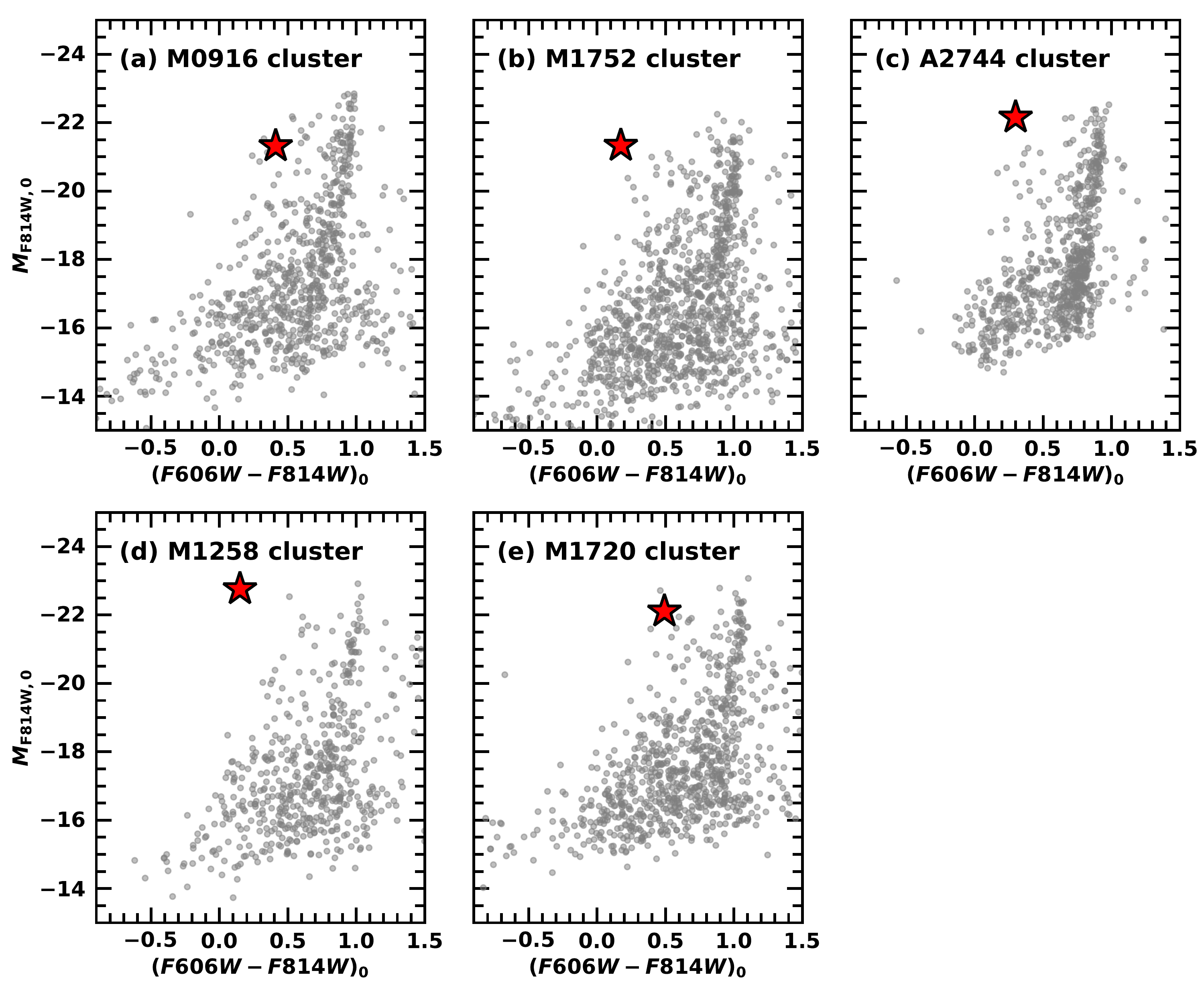}
	\caption{Color-magnitude diagrams of galaxies detected in the \textit{HST} images of the host clusters.
	All colors and magnitudes are de-reddened by the foreground extinction magnitudes.
	Red star symbols show the data of the five jellyfish galaxies.
	Gray circles show the data of all extended sources with half-light radii larger than 
	$0\farcs2$ in the angular scale and $\gtrsim1~{\rm kpc}$ in the physical scale.
	\label{fig:CMDs}}
\end{figure*}

{\color{blue} {\bf Figure \ref{fig:CMDs}}} shows the color-magnitude diagrams of galaxies in the \textit{HST} fields of the host clusters.
We ran SExtractor \citep{ber96} to detect sources in the \textit{HST} images and measure their colors and magnitudes.
For the color-magnitude diagrams, we plot the data of the extended sources (gray circles) with a half-light radius (\texttt{FLUX\_RADIUS}) larger than 4 pixels ($0\farcs2$ in the angular scale and $\gtrsim1~{\rm kpc}$ in the physical scale).
These diagrams clearly show the red sequence of cluster members in 
each \textit{HST} field.
The jellyfish galaxies in this study (red star symbols) 
have much bluer colors than the galaxies in the red sequence.
This indicates that young stellar populations are more dominant in these jellyfish galaxies than in other cluster members. 

\begin{deluxetable*}{lccccc}
	\tabletypesize{\footnotesize}
	\setlength{\tabcolsep}{0.05in}
	\tablecaption{GMOS/IFU Observations of the Jellyfish Galaxies}
	\tablehead{\colhead{Galaxy Name} & \colhead{MACSJ0916-JFG1} & \colhead{MACSJ1752-JFG2} & \colhead{A2744-F0083} & \colhead{MACSJ1258-JFG1} & \colhead{MACSJ1720-JFG1}}
	\startdata
	Program ID & GS-2019A-Q-214 & GN-2019A-Q-215 & GS-2019B-Q-219 & GN-2021A-Q-205 & GN-2021A-Q-205 \\
	Instrument & GMOS-S & GMOS-N & GMOS-S & GMOS-N & GMOS-N \\
	Observing mode & IFU two-slit & IFU two-slit & IFU two-slit & IFU two-slit & IFU two-slit \\
	Grating \& Filter & R150+GG455 & R150+GG455 & R400+CaT & R150+GG455 & R150+GG455 \\
	$\lambda_{\rm obs}~({\rm \AA})$ & $4550-9550$ & $4690-9590$ & $7820-9260$ & $4690-9280$ & $4700-9700$ \\
	$\lambda_{\rm rest}~({\rm \AA})$ & $3420-7180$ & $3470-7090$ & $6000-7110$ & $3490-6910$ & $3430-7080$ \\
	Spectral resolution\tablenotemark{\rm \footnotesize a} & 1230 & 1260 & 2440 & 1200 & 1270 \\
	Total exposure (sec) & 4320 & 15040 & 10320 & 9900 & 14400 \\ 
	Observing date (UT) & 2019-03-04 & 2019-06-11 & 2019-08-09 & 2021-05-17 & 2021-06-21 \\
	Mean airmass & 1.29 & 1.25 & 1.19 & 1.23 & 1.18 \\
	Standard star & GD108 & Wolf1346 & VMa2 & Feige34 & Feige34 \\
	Seeing FWHM\tablenotemark{\rm \footnotesize b} & 0.88\arcsec & 0.50\arcsec & 1.14\arcsec & 0.71\arcsec & 0.56\arcsec \\
	Surface brightness limit\tablenotemark{\rm \footnotesize c} & \multirow{2}*{$4.9\times10^{-20}$} & \multirow{2}*{$1.8\times10^{-19}$} & \multirow{2}*{$2.7\times10^{-19}$} & \multirow{2}*{$4.5\times10^{-19}$} & \multirow{2}*{$2.1\times10^{-20}$} \\ 
	(${\rm erg~s^{-1}~cm^{-2}~arcsec^{-2}}$) & & & & & \\
	\enddata
	\label{tab3}
	\tablenotetext{}{\textbf{Notes.}}
	\tablenotetext{\rm a}{Spectral resolution is derived from $R=\lambda/\Delta\lambda_{\rm FWHM}$ in the \Ha~emission line.}
	\tablenotetext{\rm b}{Seeing values are derived from the median FWHMs of stars detected in the following GMOS acquisition images: MACSJ0916-JFG1 ($r$-band, 60 sec), MACSJ1752-JFG2 ($r$-band, 15 sec), A2744-F0083 ($i$-band, 30 sec), MACSJ1258-JFG1 ($r$-band, 15 sec), and MACSJ1720-JFG1 ($r$-band, 15 sec).}
	\tablenotetext{\rm c}{The $3\sigma$ surface brightness level measured at the \Ha+[\ion{N}{2}] regions.}
\end{deluxetable*}

\subsection{Gemini GMOS/IFU Data}
\label{sec:gemini}

\subsubsection{Observation}
{\color{blue} {\bf Table \ref{tab3}}} summarizes our GMOS/IFU observation programs.
The five jellyfish galaxies were observed in the 2019A/B and 2021A seasons (PI: Jeong Hwan Lee): MACSJ0916-JFG1, 
MACSJ1752-JFG2, 
A2744-F0083, 
MACSJ1258-JFG1, and 
MACSJ1720-JFG1. (PID: GN-2021A-Q-205).
We obtained the IFU data of three galaxies (MACSJ1752-JFG2, MACSJ1258-JFG1, and MACSJ1720-JFG1) from the site of GMOS-North (GMOS-N) and two other targets (MACSJ0916-JFG1 and A2744-F0083) from GMOS-South (GMOS-S).
All the observations used the GMOS/IFU 2-slit mode with an  FOV of $5\arcsec\times7\arcsec$.

We observed all galaxies with the exception of A2744-F0083 
with the R150 grating and GG455 blocking filter, which securely covers a wide range of wavelengths from 4700\AA~to 9200\AA~in the observer frame ($\sim3500-6900$\AA~in the rest frame).
This coverage includes strong emission lines from [\ion{O}{2}]$\lambda\lambda3727,3729$ to [\ion{S}{2}]$\lambda\lambda6717,6731$.
We derived the spectral resolution of the spectra from the telluric emission lines obtained by the sky fibers.
The configuration of R150$+$GG455 has a low spectral resolution of $R\sim1200$ at the wavelength of the \Ha~line.
For A2744-F0083, the R400 grating and CaT filter were applied with the central wavelength of 8600\AA.
This combination had a limited wavelength coverage from 7820\AA~to 9260\AA~(6000-7110\AA~in the rest frame) but had better spectral resolution ($R\sim2400$).
As a result, the data of A2744-F0083 only covered emission lines from [\ion{O}{1}]$\lambda6300$ to [\ion{S}{2}]$\lambda\lambda6717,6731$.




MACSJ0916-JFG1 was observed 
at the  GMOS-South with an exposure time of $4\times1080~{\rm s}$ (total 1.2 hr) on the 2019-03-04 (UT).
MACSJ1752-JFG2 was observed with the exposure time of $16\times900~{\rm s}+640~{\rm s}$ (total 4.2 hr) during three nights starting from 2019-06-11 (UT).
A2744-F0083 was observed with $12\times860~{\rm s}$ (total 2.9 hr) on the UT dates from 2019-08-09 to 2019-08-30.
In the 2021A season, MACSJ1258-JFG1 and MACSJ1720-JFG1 were observed with $11\times900~{\rm s}$ (total 2.8 hr) and $16\times900~{\rm s}$ (total 4.0 hr), respectively.
The mean airmass of the data ranged from 1.2 to 1.3 for all targets.
The median full widths at half maximum (FWHMs) of the seeing in our observations were measured from the acquisition images, ranging about $0\farcs5-0\farcs7$ at the GMOS-N and $0\farcs9-1\farcs2$ at the GMOS-S.

\subsubsection{Data Reduction}
We reduced the GMOS/IFU data using the \texttt{GEMINI} package in \texttt{PyRAF}.
We used the tasks of \texttt{GBIAS}, \texttt{GFREDUCE}, and \texttt{GFRESPONSE} for bias subtraction and flat fielding with the Gemini Facility Calibration Unit (GCAL) flat.
Since there were some missing fibers that the \texttt{PyRAF/Gemini} tasks could not find, we developed our own codes\footnote[1]{\url{https://github.com/joungh93/PyRAF_GMOS_IFU}} to correct the IFU mask definition file (MDF) for proper flat fielding.
We performed the interactive tasks of \texttt{GSWAVELENGTH} to obtain the wavelength solution from the CuAr arcs.
The scattered light interfering in the fiber gaps was removed by the 2-D surface model with the order of 3 in the \texttt{GFSCATSUB} task.
After the preprocessing of the science frames, the cosmic rays and the sky lines were removed by \texttt{GEMCRSPEC} and \texttt{GFSKYSUB} tasks.
Four standard stars were used for flux calibrations of our sample: GD108 (MACSJ0916-JFG1), Wolf1346 (MACSJ1752-JFG2), VMa2 (A2744-F0083), and Feige34 (MACSJ1258-JFG1 and MACSJ1720-JFG1).
We carried out the flux calibrations by using \texttt{GSCALIBRATE} with the Chebyshev polynomials with an order of 11.
These reduction processes created individual IFU data cubes with dispersions of $1.95{\rm~\AA~pixel^{-1}}$ for the R150 grating and $0.75{\rm~\AA~pixel^{-1}}$ for the R400.
The spatial pixel scale of all cubes is $0\farcs1~{\rm pixel^{-1}}$.
We then carried out the spectral and spatial alignments of the cubes and combined them into one final cube for each target.
We used these final data cubes for the analysis of the five jellyfish galaxies in this study.

\begin{deluxetable*}{lcll}
	\tabletypesize{\footnotesize}
	\setlength{\tabcolsep}{0.20in}
	\tablecaption{Optical and Near-infrared Archival Images of the Jellyfish Galaxies}
	\tablehead{\colhead{Galaxy Name} & \colhead{Instrument} & \colhead{Filters} & \colhead{Survey\tablenotemark{\rm \footnotesize a}}}
	\startdata
	MACSJ0916-JFG1 & \textit{HST}/ACS & $F606W$, $F814W$ & PID: 10491, 12166 \\
	& \textit{WISE} & W1 (3.4 $\mu$m), W2 (4.6 $\mu$m) & All-Sky Data Release \\ \hline
	MACSJ1752-JFG2 & \textit{HST}/ACS & $F435W$, $F606W$, $F814W$ & PID: 12166, 12884, 13343 \\
	& \textit{Spitzer}/IRAC & 3.6 $\mu$m, 4.5 $\mu$m & PID: 12095 \\ \hline
	A2744-F0083 & \textit{HST}/ACS & $F435W$, $F606W$, $F814W$ & PID: 13495, 15117 \\
	& \textit{Spitzer}/IRAC & 3.6 $\mu$m, 4.5 $\mu$m & PID: 83, 90257 \\ \hline
	MACSJ1258-JFG1 & \textit{HST}/ACS & $F606W$, $F814W$ & PID: 10491, 12166 \\
	& \textit{Spitzer}/IRAC & 3.6 $\mu$m, 4.5 $\mu$m & PID: 12095 \\ \hline
	MACSJ1720-JFG1 & \textit{HST}/ACS & $F435W$, $F606W$, $F814W$ & PID: 12455 \\
	& \textit{Spitzer}/IRAC & 3.6 $\mu$m, 4.5 $\mu$m & PID: 545, 90213, 90233 \\
	\hline
	\enddata
	\label{tab4}
	\tablenotetext{}{\textbf{Note.}}
	\tablenotetext{\rm a}{\textit{HST} - PIs: H. Ebeling (PID 10491, 12166, 12884), D. Wittmann (PID 13343), B. Siana (PID 13389), J. Lotz (PID 13495), C. Steinhardt (PID 15117), and M. Postman (PID 12455), 
	\textit{Spitzer} - PIs: E. Egami (PID 545, 12095), G. Rieke (PID 83), T. Soifer (PID 90257), R. Bouwens (PID 90213), C.Lawrence (PID 90233)}
\end{deluxetable*}

\subsection{Archival Images}
\label{sec:image}

We used optical and near-infrared (NIR) images of our targets to estimate their stellar mass.
This is because the signal-to-noise ratios (S/N) of the GMOS/IFU spectra were not high enough to derive their stellar mass from spectral continuum fitting.
We collected the images from archives of the \textit{HST} for optical views and Wide-field Infrared Survey Explorer (\textit{WISE}) and \textit{Spitzer} for NIR views.
With these multi-wavelength images, we compared the morphology of the stellar emission with that of the gas emission observed by the GMOS/IFU spectral maps.
In addition, we derived stellar masses of the jellyfish galaxies using these images.
{\color{blue} {\bf Table \ref{tab4}}} lists the information of the archival images we used for analysis.

\subsubsection{\textit{HST} Optical Images}
We retrieved the \textit{HST}/ACS 
images of the five jellyfish galaxies from the Mikulski Archive for Space Telescopes (MAST).
We obtained the individual calibrated images (\texttt{*\_flc.fits} or \texttt{*\_flc.fits}) and drizzled them by using the \texttt{TweakReg} and \texttt{AstroDrizzle} tasks in \texttt{DrizzlePac}.
We combined all the images with a pixel scale of $0\farcs05~{\rm pixel^{-1}}$.

MACSJ0916-JFG1 and MACSJ1258-JFG1 were covered 
with 
two optical bands of ACS/$F606W$ and ACS/$F814W$.
Both bands were observed 
during the \textit{HST} program of the snapshot survey of the massive galaxy clusters (PI: H. Ebeling).
The drizzled \textit{HST} images had exposure times of 1200 sec for $F606W$ and 1440 sec for $F814W$.

For MACSJ1752-JFG2, we utilized 
three ACS optical bands ($F435W$, $F606W$, and $F814W$).
These images were 
obtained through 
 a snapshot survey (PI: H. Ebeling) and a program for weak lensing analysis (PI: D. Wittmann).
The exposure times of the final images were 
 2526 sec ($F435W$), 3734 sec ($F606W$), and 6273 sec ($F814W$).

A2744-F0083 was covered 
in the HFF \citep{lot17} and the Buffalo \textit{HST} survey \citep{ste20}.
We utilized 
three ACS optical bands ($F435W$, $F606W$, and $F814W$) from the surveys.
These drizzled images had the longest exposure times among our targets: 45747 sec ($F435W$), 26258 sec ($F606W$), and 108984 sec ($F814W$).

MACSJ1720-JFG1 was observed by a multi-wavelength observation program for the lensing analysis of massive clusters (PI: M. Postman).
We utilized 
three ACS optical bands ($F435W$, $F606W$, and $F814W$).
The images had exposure times of 2040 sec ($F435W$), 2020 sec ($F606W$), and 3988 sec ($F814W$), respectively.

\subsubsection{Spitzer and WISE NIR Images}


We obtained archival NIR images taken 
with \textit{Spitzer}/IRAC from the \textit{Spitzer} Heritage Archive (SHA) for all galaxies with the exception of MACSJ0916-JFG1, which was not observed by \textit{Spitzer}.
For MACSJ0916-JFG1, we retrieved the \textit{WISE} images from the NASA/IPAC Infrared Science Archive (IRSA).
The images were taken from the \textit{WISE} All-Sky Data Release and had exposure times of 7.7 sec for W1 ($3.4~{\rm \mu m}$) and W2 ($4.6~{\rm \mu m}$) bands.
As the sensitivity of the W3 ($12~{\rm \mu m}$) and W4 ($22~{\rm \mu m}$) bands are too low, 
we only used the W1 and W2 images for our analysis.
MACSJ1752-JFG2 and MACSJ1258-JFG1 were observed 
through an IRAC snapshot imaging survey of massive clusters (PID: 12095, PI: Egami) with exposure times of 94 sec ($3.6~{\rm \mu m}$) and 97 sec ($4.5~{\rm \mu m}$).
As for A2744-F0083, we combined a total of 20 IRAC exposures from two science programs: PID 83 (2 exposures; PI: Rieke) and PID 90257 (18 exposures; PI: Soifer).
The exposure times of the combined images were 1878 sec ($3.6~{\rm \mu m}$) and 1936 sec ($4.5~{\rm \mu m}$).
MACSJ1720-JFG1 had 6 IRAC exposures from three programs: PID 545 (2 exposures), PID 90213 (3 exposures), and PID 90233 (1 exposure).
After combining these images, we obtained the final images with exposure times of 374 sec ($3.6~{\rm \mu m}$) and 387 sec ($4.5~{\rm \mu m}$).

\section{Analysis}
\label{sec:analysis}



\subsection{Emission Line Fitting}
In this study, we focused on analyzing strong emission lines such as \Ha, \Hb, [\ion{O}{3}], [\ion{N}{2}], and [\ion{S}{2}] given the low S/N of the spectral continuum in the GMOS/IFU data.
First, we removed the background spectral noise from the combined IFU cubes.
We took the median value of the spaxels at the edge of the cubes, which has no object signal, as background noise in each cube.
We subtracted these noise patterns from all the pixels in the IFU cubes.
This process cleaned the unnecessary instrumental noise pattern of the spectra.
Next, we carried out Voronoi binning with the Python \texttt{vorbin} package \citep{cap03} to obtain high S/N for emission line analysis in each bin.
The S/N in the Voronoi bins ranged from 30 to 60 at the \Ha+[\ion{N}{2}] regions for the jellyfish galaxies, depending on the exposure time and the quality of the data.
The numbers of the final Voronoi bins in the IFU cubes were 42 (MACSJ0916-JFG1), 147 (MACSJ1752-JFG2), 175 (A2744-F0083), 215 (MACSJ1258-JFG1), and 166 (MACSJ1720-JFG1).
Then, we subtracted the continuum from the spectra of all the Voronoi bins.
We performed the Gaussian smoothing for continuum subtraction because it was difficult to carry out the spectral continuum fitting (e.g., pPXF) for the Voronoi bins with low S/N in the outer region.
For the Gaussian smoothing of the continuum, we applied a kernel width of 10\AA~after masking the emission lines.

We then fitted strong emission lines (\Hb, [\ion{O}{3}]$\lambda\lambda4959,5007$, \Ha, [\ion{N}{2}]$\lambda\lambda6548,6584$, and [\ion{S}{2}]$\lambda\lambda6717,6731$) with the Markov chain Monte Carlo (MCMC) method using the Python \texttt{emcee} package \citep{for13}.
For A2744-F0083, we fitted only the \Ha+[\ion{N}{2}] doublet and [\ion{S}{2}] doublet lines.
We applied multiple Gaussian functions for fitting the line profiles.
For the narrow components of the emission lines, we used three Gaussian profiles for the \Hb+[\ion{O}{3}] and \Ha+[\ion{N}{2}] regions and two for the [\ion{S}{2}] doublet.
In the case of AGN host galaxies (A2744-F0083 and MACSJ1258-JFG1), more profiles were added for fitting the broad components of the \Hb+[\ion{O}{3}] and \Ha+[\ion{N}{2}] regions.

We initially fitted the emission lines from the central Voronoi bins with high S/N.
We then used the initial solutions to run the MCMC process for all the bins.
We defined the prior distributions of parameters as Gaussian functions centered on the initial solutions.
We implemented the MCMC process in \texttt{emcee} with 50 walkers and 2000 steps.
After finishing all the MCMC steps of walkers, we measured the skewness and kurtosis of the posterior distributions of parameters to check the reliability of our parameter estimation.
We rejected solutions from Voronoi bins with values of skewness and kurtosis higher than 0.5 or lower than $-0.5$.
We took the median values of the posterior distributions as the burn-in solutions of the MCMC process.

\begin{figure*}
	\centering
	\includegraphics[width=\textwidth]{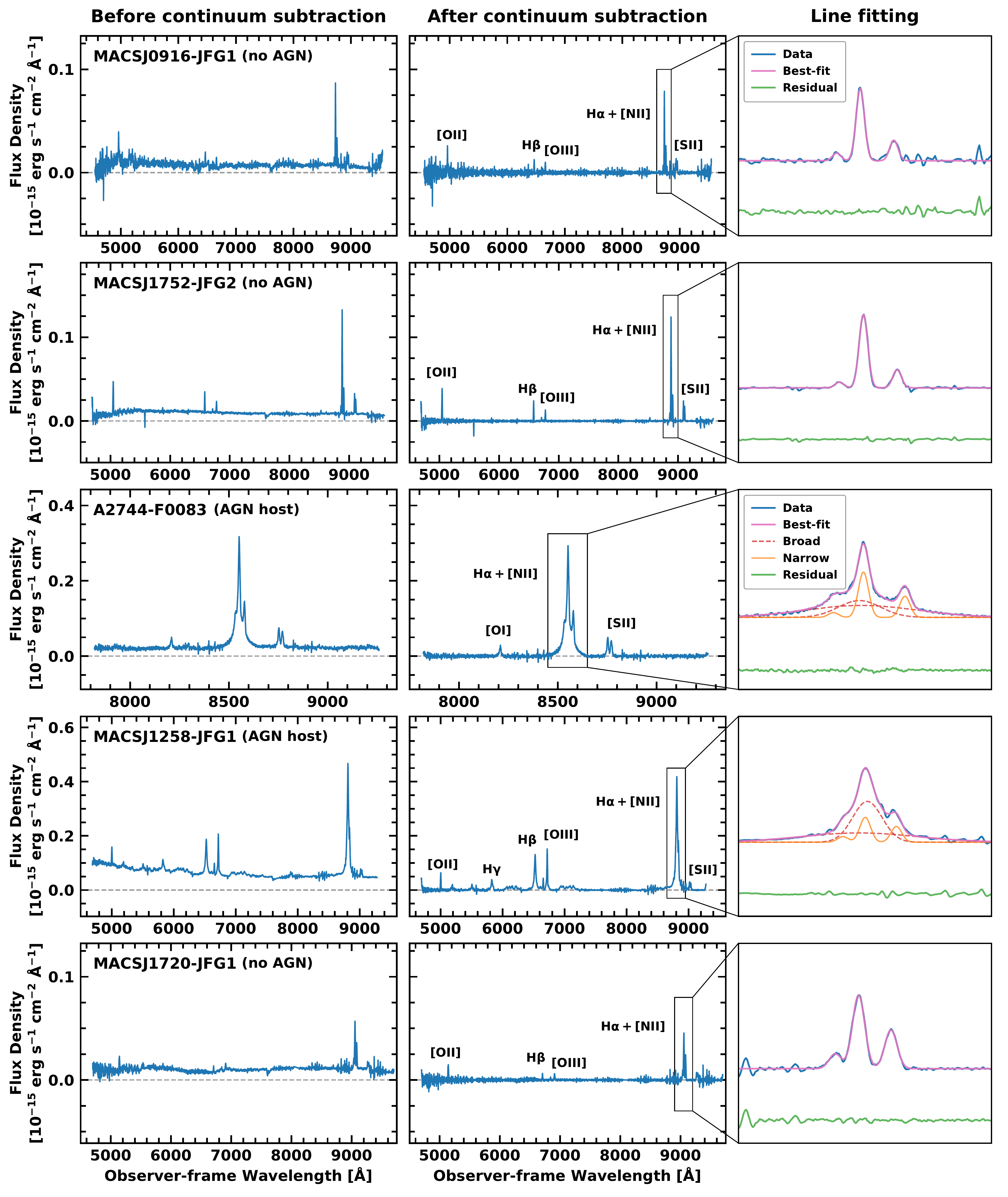}
	\caption{
	The integrated spectra within a radius of within a radius of $0\farcs4$ from the galactic center of jellyfish galaxies in this study.
	The left panels show the original spectra before continuum subtraction, and the middle panels show the continuum subtracted spectra.
	In the middle panels, several strong emission lines are marked in each spectrum.
	The right zoom-in panels show the 
	${\rm H\alpha}$+[\ion{N}{2}] region of each galaxy.
	We plot the observed spectra (blue solid lines), our best-fit models (pink solid lines), and residuals (green solid lines).
	In the case of AGN host galaxies, we plot narrow and broad components of the emission lines with orange solid lines and red dashed lines, respectively.
	\label{fig:spec}}
\end{figure*}

In {\color{blue} {\bf Figure \ref{fig:spec}}}, we plot the GMOS/IFU spectra in the central regions (within a radius of $0\farcs$4) of our targets.
We used these integrated spectra for obtaining the initial solutions of the line fitting in all the Voronoi bins.
We plot the original spectra and the continuum subtracted spectra in the left and middle panels, respectively.
The right panels show the zoom-in spectra at the \Ha+[\ion{N}{2}] regions overlaid with the best-fit models (pink solid lines).
For the AGN host galaxies, 
we added two more profiles (red dashed lines) for broad components of this region.
Residuals (green solid lines) show that the best-fit models represent the spectra well.






\subsection{Measurements of SFRs, Gas Velocity Dispersion, and Stellar Masses}

\subsubsection{$H\alpha$-derived SFRs}
\label{sec:sfr}

In this section, we describe how we estimated the SFRs based on the emission line analysis.
We derived the SFRs of jellyfish galaxies from their \Ha~luminosity that was corrected for stellar absorption, extinction, and AGN contamination.

Stellar absorption affects the observed fluxes of the Balmer lines such as \Ha~and \Hb.
Since the S/N of stellar continuum in our GMOS/IFU data is too low to do the full spectrum fitting, 
we took a simple approach for absorption correction by assuming constant equivalent widths (EWs) of the Balmer absorption lines \citep{hop03, gun11}.
We used the following formula for the absorption correction,
\begin{equation}
    S_{\rm H\alpha}=F_{\rm H\alpha}\times\frac{{\rm EW_{H\alpha}}+{\rm EW}_{c}}{\rm EW_{H\alpha}},
\end{equation}
where $S_{\rm H\alpha}$ is the absorption-corrected \Ha~flux, $F_{\rm H\alpha}$ is the observed \Ha~flux, ${\rm EW_{H\alpha}}$ is the observed \Ha~EW, and ${\rm EW}_{c}$ is the EW for stellar absorption.
\citet{hop03} suggested that the median intrinsic value of ${\rm EW}_{c}$ is 2.6\AA~for the SDSS sample of star-forming galaxies but this value would be reduced to 1.3\AA~in the real SDSS spectra.
This is because SDSS spectra have sufficient spectral resolution ($R\sim2000$) to resolve broad stellar absorption profiles (see Figure 7 in \citet{hop03}).
Therefore, we adopted ${\rm EW}_{c}=1.3{\rm \AA}$ for A2744-F0083 observed by the R400 grating with a similar resolution ($R\sim2400$) to the SDSS and ${\rm EW}_{c}=2.6{\rm \AA}$ for the remaining targets observed with the R150 grating ($R\sim1200$).
We applied the same EWs to correct the absorption of \Hb~line.

We subsequently corrected for internal dust extinction and foreground extinction.
For the dust extinction correction, we used the absorption-corrected flux ratio of the \Ha~and \Hb~lines, otherwise known as the Balmer decrement.
We assumed an intrinsic \Ha~to \Hb~flux ratio of 2.86 
for the case B recombination with $T_{e}=10000~{\rm K}$ and adopted the \citet{car89} extinction law.
The equations for extinction correction are given as below,
\begin{equation}
    A_{\rm H\alpha}=k_{\rm H\alpha}\times E(B-V),
\end{equation}
where $A_{\rm H\alpha}$ is the extinction magnitude of the \Ha~flux, $k_{\rm H\alpha}$ is the extinction coefficient for \Ha, and $E(B-V)$ is the color excess.
We adopted $k_{\rm H\alpha}=2.53$ from \citet{car89}.
The color excess can be derived from the equation (\ref{eqn:ebv}).
\begin{equation}
    \label{eqn:ebv}
    E(B-V)=2.33\times\log\left[\frac{S_{\rm H\alpha}/S_{\rm H\beta}}{2.86}\right],
\end{equation}
where $S_{\rm H\alpha}/S_{\rm H\beta}$ is the absorption-corrected \Ha~to \Hb~flux ratio.
We also corrected for foreground extinction using the reddening magnitudes from \citet{sch11} (see {\color{blue} {\bf Table \ref{tab2}}}).

From the corrected \Ha~flux, we derived the SFRs using the following relation given by \citet{ken98} assuming the \citet{cha03} initial mass function (IMF),
\begin{equation}
    \label{eqn:SFR}
    {\rm SFR_{H\alpha}}=4.6\times10^{-42}L_{\rm H\alpha},
\end{equation}
where ${\rm SFR_{H\alpha}}$ is the \Ha-based SFR with unit of $M_{\odot}~{\rm yr^{-1}}$ and $L_{\rm H\alpha}$ is the \Ha~luminosity corrected for stellar absorption and extinction.
The extinction law and IMF adopted in this paper are consistent with those in the GASP studies.

Throughout these processes, we rejected the Voronoi bins with S/N (\Ha) lower than 3.
In addition, we excluded the Voronoi bins classified as AGN and low-ionization nuclear emission-line regions (LINERs) in the BPT diagram, as shown in {\color{blue} {\bf Figures \ref{fig:BPT1}}} and {\color{blue} {\bf \ref{fig:BPT2}}} (see {\color{blue} {\bf Section \ref{sec:bpt}}}).
For the Voronoi bins with $\rm S/N<3$ in \Hb~and [\ion{O}{3}], we used the [\ion{N}{2}]$\lambda6584/{\rm H\alpha}$ ratio for the classification instead.
We classified the Voronoi bins with log([\ion{N}{2}]$\lambda6584/{\rm H\alpha})>-0.4$ as AGN and excluded them from consideration.
We also included the Voronoi bins with ${\rm S/N~(H\alpha)>3}$ but weak S/N ($<3$) of \Hb~and the BPT forbidden lines ([\ion{\rm O}{3}] and [\ion{\rm N}{2}]) for computing the SFR, as done in \citet{medl18}.

As a result, total SFR values in each correction step (continuum subtraction, absorption correction, and extinction correction) for the five jellyfish galaxies are listed in {\color{blue} {\bf Table \ref{tab5}}}.
This table shows that dust extinction correction takes much more effect on the SFR values than the two other correction steps.

\begin{deluxetable*}{ccccc}
	\tabletypesize{\footnotesize}
	\setlength{\tabcolsep}{0.10in}
	\tablecaption{
	Total SFR Calculations of the Jellyfish Galaxies in Each Step}
	\tablehead{\colhead{Galaxy Name} & \colhead{Initial SFR} & \colhead{Continuum subtraction} & \colhead{Absorption correction} & \colhead{Extinction correction} \\
	 & ($M_{\odot}~{\rm yr^{-1}}$) & ($M_{\odot}~{\rm yr^{-1}}$) & ($M_{\odot}~{\rm yr^{-1}}$) & ($M_{\odot}~{\rm yr^{-1}}$) \\
	(1) & (2) & (3) & (4) & (5)}
	\startdata
	MACSJ0916-JFG1 & 7.21 & 5.47 & 5.64 & \textbf{7.12} \\
	MACSJ1752-JFG2 & 11.43 & 8.79 & 9.03 & \textbf{30.43} \\
	A2744-F0083$^{\ast}$ & 12.76 & 9.08 & 9.17 & \textbf{22.00} \\
	MACSJ1258-JFG1$^{\ast}$ & 17.89 & 10.37 & 11.37 & \textbf{35.71} \\
	MACSJ1720-JFG1 & 13.75 & 7.14 & 7.62 & \textbf{17.37} \\
	\enddata
	\label{tab5}
	\tablenotetext{}{\textbf{Notes.}}
	\tablenotetext{}{Columns are: (1) Name of jellyfish galaxy; (2) Total SFR values from the original spectra; (3) Total SFR values from the spectra after continuum subtraction; (4) Total SFR values from the spectra after continuum subtraction and stellar absorption correction; (5) Final SFR values from the spectra after continuum subtraction, stellar absorption correction, and dust extinction correction.}
	\tablenotetext{}{Asterisks ($^{\ast}$) mark  
	AGN host galaxies.}	
\end{deluxetable*}

\subsubsection{Gas Velocity Dispersion}
\label{sec:vdgas}

We derived the gas velocity dispersion ($\sigma_{v, {\rm gas}}$) using the standard deviation of \Ha~emission line corrected for beam-smearing and instrumental dispersion.
Beam-smearing occurs when the observed line in a IFU spaxel is blended with the lines from the neighboring spaxels, which makes a broader line profile.
Since the seeing FWHM in our observations ranged from $0\farcs5$ to $1\farcs1$, which corresponds from 5 to 11 spaxels for the GMOS/IFU, this effect significantly contributes to the velocity dispersion.
To correct for beam-smearing, we followed the method described in Appendix A in \citet{sto16}.
Using the artificial VLT/KMOS data cube with similar seeing and pixel scale to our GMOS/IFU data, they suggested that a linear subtraction of velocity gradient ($\Delta v/\Delta r$) from the observed velocity dispersion can effectively correct for beam-smearing.
This method is described with the equation below,
\begin{equation}
    \sigma_{v,{\rm cor}}=\sigma_{v,{\rm obs}}-\frac{\Delta v}{\Delta r},
\end{equation}
where $\sigma_{v,{\rm cor}}$ is the beam-smearing corrected velocity dispersion, $\sigma_{v,{\rm obs}}$ is the observed velocity dispersion, and $\Delta v/\Delta r$ is the maximum velocity gradient at a distance equivalent to the seeing FWHM from the center of a galaxy.
Our data cubes showed maximum velocity gradients of $\Delta v/\Delta r=4.9~{\rm km~s^{-1}~spaxel^{-1}}$ (MACSJ0916-JFG1), $8.2~{\rm km~s^{-1}~spaxel^{-1}}$ (MACSJ1752-JFG2), $10.6~{\rm km~s^{-1}~spaxel^{-1}}$ (A2744-F0083), $10.5~{\rm km~s^{-1}~spaxel^{-1}}$ (MACSJ1258-JFG1), and $14.8~{\rm km~s^{-1}~spaxel^{-1}}$ (MACSJ1720-JFG1).
For comparison, \citet{sto16} measured $\Delta v/\Delta r=13.4~{\rm km~s^{-1}~spaxel^{-1}}$.

Then, we subtracted the GMOS/IFU instrumental dispersion from the beam-smearing corrected dispersion.
The instrumental velocity dispersion was derived from a single Gaussian fitting of the sky emission lines which were extracted during the data reduction process.
We obtained the instrumental velocity dispersion at the wavelength of \Ha~for our GMOS/IFU data: $\sigma_{v,{\rm inst}}=103.1~{\rm km~s^{-1}}$ (MACSJ0916-JFG1), $101.4~{\rm km~s^{-1}}$ (MACSJ1752-JFG2), $52.2~{\rm km~s^{-1}}$ (A2744-F0083), $106.1~{\rm km~s^{-1}}$ (MACSJ1258-JFG1), and $100.1~{\rm km~s^{-1}}$ (MACSJ1720-JFG1).
We used a quadrature removal of instrumental dispersion,
\begin{equation}
    \sigma_{v,{\rm gas}}^{2}=\sigma_{v,{\rm cor}}^{2}-\sigma_{v,{\rm inst}}^{2},
\end{equation}
where $\sigma_{v,{\rm gas}}$ is the corrected gas velocity dispersion and $\sigma_{v,{\rm inst}}$ is the instrumental velocity dispersion.

\subsubsection{Stellar Masses}

We estimated the stellar masses ($M_{\ast}$) of our jellyfish galaxies using the NIR images described in {\color{blue} {\bf Section \ref{sec:image}}}.
The GASP studies derived the stellar masses from the spectral continuum fitting \citep{pog17}.
However, our GMOS/IFU data showed too low S/N of the continuum to apply the same method as the GASP studies.
Instead, we converted the NIR fluxes of the \textit{Spitzer}/IRAC $3.6~{\rm \mu m}$ and $4.5~{\rm \mu m}$ (W1 and W2 in the case of \textit{WISE} data) to stellar masses following the relation from \citet{esk12}.
\begin{equation}
    M_{\ast}=\xi\times10^{5.65}\times S^{2.85}_{\rm 3.6\mu m}S^{-1.85}_{\rm 4.5\mu m}\times(d_{L}/0.05)^{2}~M_{\odot},
\end{equation}
where $\xi$ is the conversion factor from the Salpeter IMF \citep{sal55} to other IMFs, $S_{\rm 3.6\mu m}$ and $S_{\rm 4.5\mu m}$ are the NIR fluxes of $3.6~{\rm \mu m}$ and $4.5~{\rm \mu m}$ in Jy, and $d_{L}$ is the luminosity distance in Mpc.
We applied $\xi=0.54$ for the Chabrier IMF \citep{cha03} with a mass range from 0.1 to 100 $M_{\odot}$.
This method has been used for estimating the stellar masses of star-forming galaxies in Abell 2744 \citep{raw14}.

We utilized \texttt{photutils v1.0.2} \citep{bra21} to measure the NIR fluxes of the jellyfish galaxies.
We performed aperture photometry using rectangular apertures with the size of the GMOS/IFU FOV ($5''\times7''$) for all targets except for MACSJ0916-JFG1.
Since MACSJ0916-JFG1 was only observed by \textit{WISE} with poor spatial resolution \citep[$\rm FWHM\sim6\farcs4$ in W1 and W2;][]{wri10}, we increased the aperture size to twice the FWHM of the \textit{WISE} images.
The derived stellar masses are listed in {\color{blue} {\bf Table \ref{tab6}}}, ranging from ${\rm log}~M_{\ast}/M_{\odot}=9.84$ to 10.90.

\subsection{Definition of Disk and Tail}
\label{sec:bound}

In this study, we spatially divided a jellyfish galaxy into two components -- ``disk'' and ``tail''.
The disk corresponds to the main body of the galaxy, 
and the tail is the extraplanar region composed of ram-pressure stripped ISM outside the disk.
To compare our results with the GASP studies, we used the definition of the galaxy boundary between the disk and tail as described in \citet{pog19} and \citet{gul20}.
We made use of the emission line maps of the \Ha+[\ion{N}{2}] region from the GMOS/IFU data.
Although the GASP studies used continuum maps of the \Ha+[\ion{N}{2}] region, it was difficult for this study to use continuum maps due to low S/N.
The background level and standard deviation ($1\sigma$) were estimated using the DAOPHOT MMM algorithm on the $3\sigma$-clipped data in the emission line maps.
We then manually masked the disturbed region of the galaxy, removing tails and extraplanar clumps.
With the masked map, we carried out elliptical isophote fitting using \texttt{PyRAF/Ellipse}.
The galaxy boundary between the disk and tail was defined as an elliptical isophote with a surface brightness of $1\sigma$ above the background level.
The upper right panels of {\color{blue} {\bf Figures \ref{fig:IFU1}-\ref{fig:IFU5}}} display the defined boundaries in the five jellyfish galaxies with red dashed ellipses.

Although we followed the definition from the GASP studies, the boundary between the disk and tail used in this paper could not be fully consistent with that in the GASP studies.
This is because manual masking of disturbed regions is a little arbitrary and the spatial resolution of the GMOS/IFU data is quite different from that of MUSE data in the physical scale (e.g., the mean seeing FWHM $0\farcs7$ corresponds to $>3~{\rm kpc}$ for our targets but $\sim0.7~{\rm kpc}$ in the GASP targets).
These limitations could contribute to 
large fluctuations of the fraction of SFR in tails as shown in {\color{blue} {\bf Table \ref{tab6}}}.

\begin{figure}[h]
	\centering
	\includegraphics[width=0.5\textwidth]{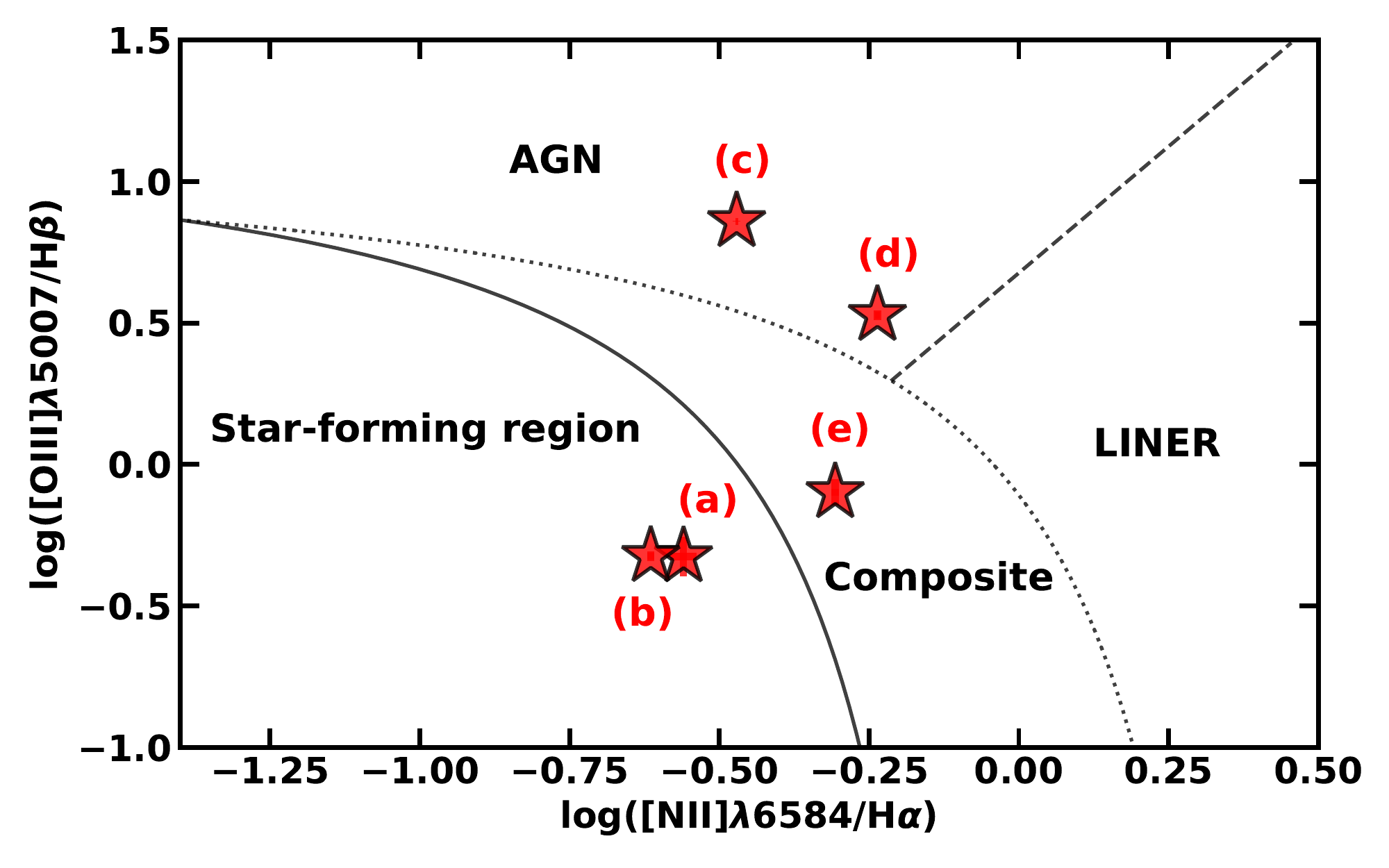}
	\caption{
	A BPT diagram of the integrated spectra of the five jellyfish galaxies.
	Each galaxy is marked with the same order 
	as in {\color{blue} {\bf Figure \ref{fig:HST2}}}: (a) MACSJ0916-JFG1, (b) MACSJ1752-JFG2, (c) A2744-F0083, (d) MACSJ1258-JFG1, and (e) MACSJ1720-JFG1.
	We plot a solid line from \citet{kau03}, a dotted line from \citet{kew01}, and a dashed line from \citet{sha10} to divide star-forming, composite, AGN, and LINER regions in this diagram.
	\label{fig:BPT1}}
\end{figure}

\begin{figure*}
	\centering
	\includegraphics[width=\textwidth]{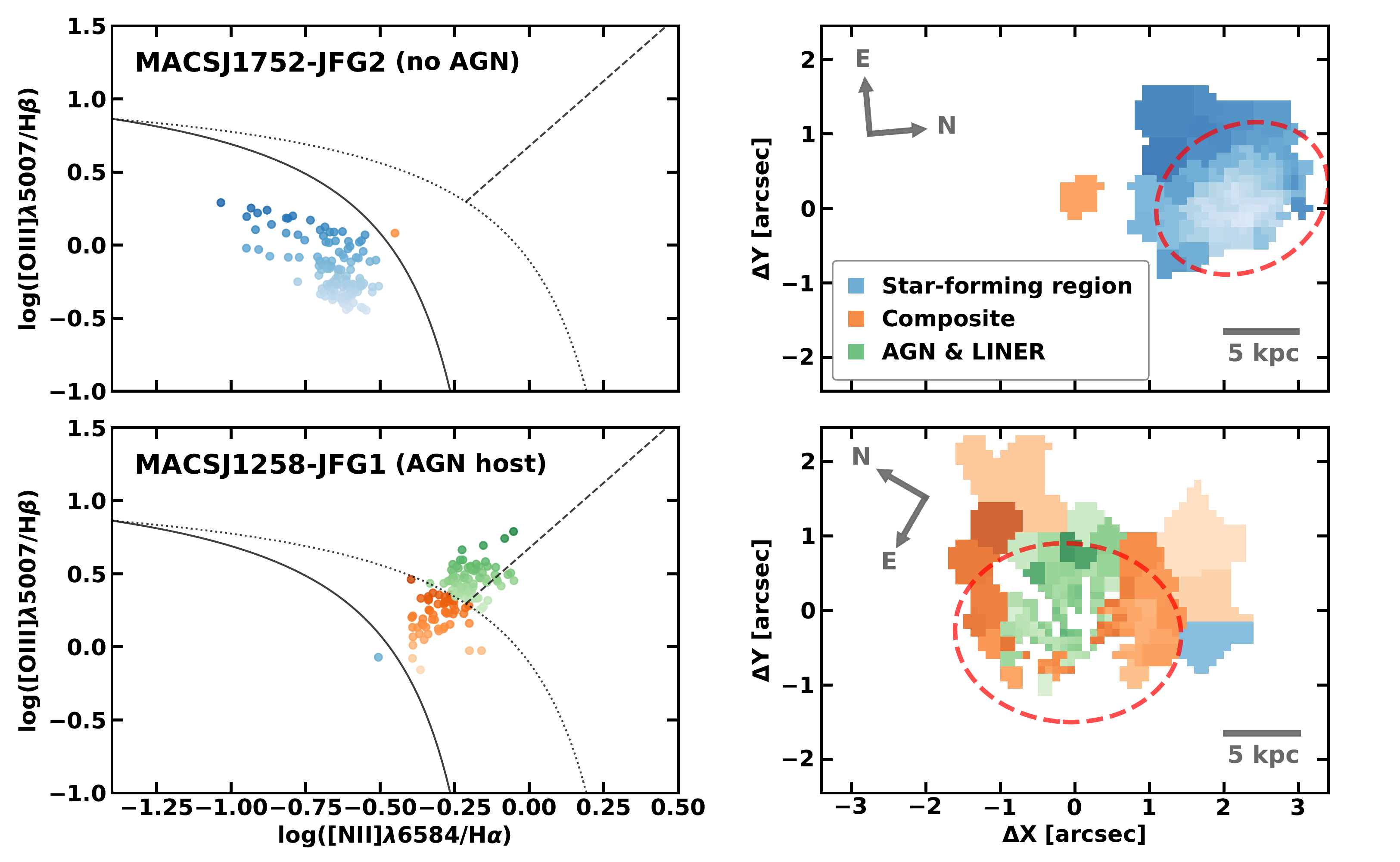}
	\caption{
	\textit{Left panels:} BPT diagrams of the Voronoi bins with ${\rm S/N}>3$ in MACSJ1752-JFG2 (upper: a non-AGN host galaxy) and MACSJ1258-JFG1 (lower: an AGN host galaxy).
	The colors of the circle symbols represent the BPT classifications: star-forming (blue), composite (orange), and AGN$+$LINER regions (green).
	We color-code each symbol with the value of log ([\ion{\rm O}{3}] 
	$\lambda5007/{\rm H}\beta)$.
	\textit{Right panels:} Spatial maps of the regions classified in the BPT diagrams.
	Here we exclude the Voronoi bins with  S/N lower than 3.
	Red dashed ellipse represents the boundary of the disk and tail of this galaxy.
	The orientation and distance scale are marked in the figure.
	\label{fig:BPT2}}
\end{figure*}

\section{Ionized Gas Properties of the Jellyfish Galaxies}
\label{sec:result}

\subsection{BPT Diagnostics}
\label{sec:bpt}

In {\color{blue} {\bf Figure \ref{fig:BPT1}}}, we display the BPT diagram ([\ion{O}{3}]$\lambda$5007/\Hb~vs. [\ion{N}{2}]$\lambda$6584/\Ha) of the five jellyfish galaxies using the narrow components of the emission lines from their integrated spectra.
All data used except for A2744-F0083 were taken with the GMOS/IFU.
For the BPT diagram of A2744-F0083, we derived their flux ratios from the AAOmega spectra \citep{owe12} because the GMOS/IFU spectra of this galaxy did not cover the \Hb~and [\ion{O}{3}] region.
To determine the gas ionization mechanisms of each galaxy, we classify galaxies into four regions (star-forming, composite, AGN, and LINER) by adopting the well-known boundaries from \citet{kew01}, \citet{kau03}, and \citet{sha10}, as done in the GASP studies.

MACSJ0916-JFG1 (a) and MACSJ1752-JFG1 (b) are located in the star-forming region, showing the lowest flux ratios of log([\ion{O}{3}]$\lambda5007/{\rm H\beta})<-0.3$ and log([\ion{N}{2}]$\lambda6584/{\rm H\alpha})<-0.6$  
among our sample.
A2744-F0083 (c) and MACSJ1258-JFG1 (d) are genuine AGN host galaxies with log([\ion{O}{3}]$\lambda5007/{\rm H\beta})>0.5$. 
These galaxies also show broad components of the Balmer emission lines (\Ha~and \Hb) in their spectra, indicating that they are type I AGNs.
MACSJ1720-JFG1 (e) is located in the composite region, showing a higher ratio of log([\ion{N}{2}]$\lambda6584/{\rm H\alpha})\sim-0.3$ than those from the two star-forming galaxies (log([\ion{N}{2}]$\lambda6584/{\rm H\alpha})\sim-0.6$).
This indicates that the gas ionization in MACSJ1720-JFG1 cannot be solely explained by star formation.
However, it is difficult to say that this galaxy hosts a type I AGN in its center, considering that its Balmer emission lines did not show any broad components in our GMOS/IFU spectra and were fitted well with single Gaussian components.
It seems that other excitation mechanisms such as shocks \citep{ric11} or heat conduction \citep{bos16} could be responsible for ionizing photons in this galaxy.

{\color{blue} {\bf Figure \ref{fig:BPT2}}} shows the BPT diagrams of Voronoi bins (left) and their spatial maps (right) of the two jellyfish galaxies, which have relatively high S/N of the BPT emission lines, as examples of a non-AGN host galaxy (MACSJ1752-JFG2) and an AGN host galaxy (MACSJ1258-JFG1). 
We only plot the bins with S/N higher than 3 for all the BPT emission lines.
In the spatial map, we display all bins with BPT classifications of star-forming (blue), composite (orange), and AGN+LINER (green).
For the remaining three galaxies, the Voronoi bins of the GMOS/IFU data show very low S/N of \Hb~and [\ion{O}{3}] lines (MACSJ0916-JFG1 and MACSJ1720-JFG1) or do not cover the wavelength range of the \Hb~and [\ion{O}{3}] lines (A2744-F0083).
We also plot the boundary between the disk and tail of each galaxy (red dashed ellipse) as obtained from the method described in {\color{blue} {\bf Section \ref{sec:bound}}}.

For MACSJ1752-JFG2, almost all bins belong to the star-forming region except for one bin in the tail region.
The central region shows log([\ion{N}{2}]$\lambda6584/{\rm H\alpha})\sim-0.6$ and log([\ion{O}{3}]$\lambda5007/{\rm H\beta})\sim-0.3$, which are consistent with typical star-forming galaxies.
It is also remarkable that the eastern tail region exhibits a higher [\ion{O}{3}]/\Hb~ratio (log([\ion{O}{3}]$\lambda5007/{\rm H\beta})>0$) and a lower [\ion{N}{2}]/\Ha~ratio (log([\ion{N}{2}]$\lambda6584/{\rm H\alpha})<-0.75$) than the central region.
The ``composite'' bin in the southern tail has marginally low S/N, such that the gas ionization source is not clear.

MACSJ1258-JFG1 has high flux ratios of log([\ion{O}{3}]$\lambda5007/{\rm H\beta})\sim 0.5$ and log([\ion{N}{2}]$\lambda6584/{\rm H\alpha})\sim-0.25$, indicating that it hosts an AGN.
In this figure, several Voronoi bins in the central region and the head region in the east show low S/N of \Hb~and [\ion{O}{3}] lines.
However, the Voronoi bins with ${\rm S/N}>3$ show that there is a clear radial gradient of the BPT line ratios from the center to the outer region.
The outer disk and tail regions on the western side are classified as composite regions, implying that both star formation and shock-heating mechanisms contribute to gas ionization there.

\subsection{SFRs, Gas Kinematics, and Dynamical States}
\label{sec:ion}

\begin{figure*}
	\centering
	\includegraphics[width=\textwidth]{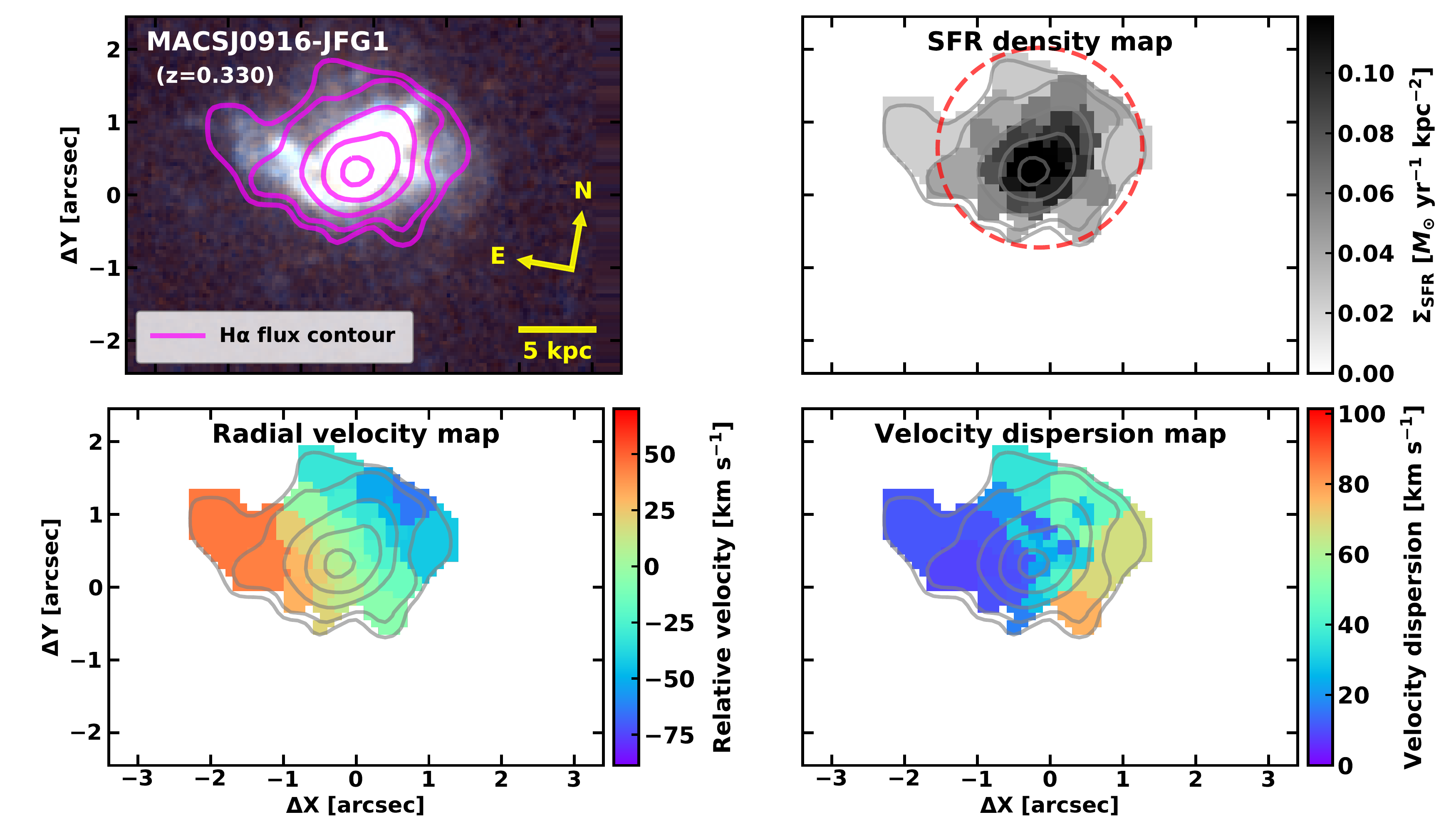}
	\caption{
	\textit{Upper left:} \textit{HST} color image of MACSJ0916-JFG1 overlaid with the \Ha~flux contour (magenta solid lines) derived from the GMOS/IFU data.
	We mark the orientation, distance scale, and redshift of this galaxy.
	\textit{Upper right:} The SFR density (${\rm \Sigma_{SFR}}$) map estimated 
	from the \Ha~luminosity.
	Red dashed circle represents the boundary of the disk and tail of this galaxy.
	Gray contours denote the spatial distribution of \Ha~flux.
	Right color bar shows the scale and range of the values.
	\textit{Lower left:} The radial velocity map derived from the peak wavelength of \Ha~emission line.
	We display the relative velocity of each bin with respect to the central Voronoi bin with the highest \Ha~flux.
	\textit{Lower right:} The gas velocity dispersion map derived from the width of \Ha~emission line.
	\label{fig:IFU1}}
\end{figure*}

We mainly analyze the ionized gas properties of each jellyfish galaxy using \Ha~emission lines.
From {\color{blue} {\bf Figures \ref{fig:IFU1}}} to {\color{blue} {\bf \ref{fig:IFU5}}}, we map the distributions of the \Ha~flux, the SFR surface density ($\Sigma_{\rm SFR}$), the radial velocity, and the gas velocity dispersion.
We select the Voronoi bins with ${\rm S/N~(H\alpha)} > 3$ for the analysis.
{\color{blue} {\bf Table \ref{tab6}}} lists the stellar masses, dust extinction, SFRs, tail SFR fraction of total SFR ($f_{\rm SFR}={\rm SFR(tail)/SFR(total)}$), and gas kinematics of the jellyfish galaxies.

To compute and correct the SFRs, we derive dust extinction magnitudes from the flux-weighted mean Balmer decrement (${\rm H\alpha/H\beta}$).
While the Balmer decrement and the extinction magnitude can slightly decrease from the center to the outskirts (a median decrease of $\sim0.25~{\rm mag}$ reported in \citet{pog17}), we choose to assume a constant value throughout the galaxy due to the poor S/N of the \Hb~emission lines in the outer regions.
As a result, the five jellyfish galaxies in this study shows a median total SFR of 22.0 \Moyr~and a median tail SFR of 6.8 \Moyr.
These SFRs are much higher than those of the GASP sample, with a median total SFR of 1.1 \Moyr~and a median tail SFR of 0.03 \Moyr.
In addition, the median SFR fraction in the tail is also much higher in this study ($f_{\rm SFR}=22\%$) than in the GASP studies ($f_{\rm SFR}=3\%$).
\citet{lee22} presented a detailed comparison of the star formation activity of the five jellyfish galaxies in this study  with that of other known jellyfish galaxies including the GASP sample, 
considering galaxy stellar mass, redshift, and jellyfish morphology.
As a result, they suggested a positive correlation between the star formation activity of jellyfish galaxies and the host cluster properties.

For ionized gas kinematics, we derive radial velocities using the peak wavelength measured from fitting.
We show line-of-sight radial velocity maps measured with respect to the central Voronoi bins.
Following the definition used in \citet{wis15}, We calculate the maximum rotational velocity at a circularized radius of the disk ($R_{\rm disk,c}=\sqrt{ab}$, where $a$ and $b$ are the galactocentric radii of the semi-major and semi-minor axis) as an indicator of the rotational speed of ionized gas within the disk.

In addition, we derive gas velocity dispersions as described in {\color{blue} {\bf Section \ref{sec:vdgas}}} to investigate the dynamical states of the ionized gas in each jellyfish galaxy. 
The median \Ha~velocity dispersion obtained by the GASP studies is $27~{\rm km~s^{-1}}$, suggesting that the \Ha-emitting clumps are dynamically cold \citep{bel19, pog19}.
With this information, we compare the distribution of the gas velocity dispersions in our sample with that of the GASP jellyfish galaxies.

\begin{deluxetable*}{ccccccccc}
	\tabletypesize{\footnotesize}
	\setlength{\tabcolsep}{0.07in}
	\tablecaption{
	Properties of the Jellyfish Galaxies}
	\tablehead{\colhead{Galaxy Name} & \colhead{$\log M_{\ast}$} & \colhead{${\rm H\alpha/H\beta}$} & \colhead{$A_{V}$} & \colhead{Total SFR} & \colhead{Tail SFR} & \colhead{$f_{\rm SFR}$} & \colhead{$|\Delta v|_{\rm disk}$} & \colhead{$\sigma_{v,{\rm gas}}$} \\
	& ($M_{\odot}$) & & (mag) & ($M_{\odot}~{\rm yr^{-1}}$) & ($M_{\odot}~{\rm yr^{-1}}$) & (\%) & (${\rm km~s^{-1}}$) & (${\rm km~s^{-1}}$) \\
	(1) & (2) & (3) & (4) & (5) & (6) & (7) & (8) & (9)}
	\startdata
	MACSJ0916-JFG1 & 10.18 & $3.16\pm0.29$ & 0.31 & $7.12\pm0.36$ & $0.56\pm0.26$ & 7.9 & 54.4 & $30.0\pm13.0$ \\
	MACSJ1752-JFG2 & 9.84 & $4.78\pm0.10$ & 1.61 & $30.43\pm0.17$ & $6.76\pm0.13$ & 22.2 & 73.2 & $43.1\pm12.6$ \\
	A2744-F0083$^{\ast}$ & 10.59 & $4.14\pm0.06$ & 1.16 & $22.00\pm0.10$ & $7.53\pm0.08$ & 34.2 & 89.5 & $45.3\pm5.3$ \\
	MACSJ1258-JFG1$^{\ast}$ & 10.90 & $4.64\pm0.19$ & 1.52 & $35.71\pm0.25$ & $16.80\pm0.20$ & 47.0 & 134.5 & $100.5\pm15.0$ \\
	MACSJ1720-JFG1 & 10.64 & $4.05\pm1.40$ & 1.09 & $17.37\pm1.12$ & $3.28\pm0.81$ & 18.9 & 103.5 &  $104.5\pm15.8$ \\
	\enddata
	\label{tab6}
	\tablenotetext{}{\textbf{Notes.}}
	\tablenotetext{}{Columns are: (1) Name of jellyfish galaxy; (2) NIR-derived stellar mass within the GMOS/IFU FOV; (3) Mean flux ratio of \Ha~to \Hb~derived from the integrated spectra; (4) The $V$-band magnitude of dust extinction derived from (3); (5) Total SFR value estimated when (4) is uniformly applied; (6) SFR in the tail region; (7) The tail SFR fraction of total SFR ($f_{\rm SFR}={\rm SFR(tail)/SFR(total)}$); (8) The maximum rotational velocity at a radius of disk; (9) Flux-weighted mean value of gas velocity dispersion.}
	\tablenotetext{}{Asterisks ($^{\ast}$) mark  
	AGN host galaxies.}	
\end{deluxetable*}

\subsubsection{MACSJ0916-JFG1}
In {\color{blue} {\bf Figure \ref{fig:IFU1}}}, it is clear that the \Ha~flux distribution of MACSJ0916-JFG1 is well-matched with its optical light distribution.
The asymmetric feature of the \Ha~flux perfectly reproduces the eastern tail in the optical image.
The \Ha~flux contours confirm that recent star formation has occurred in the blue knots in the disk and tail.

In the SFR surface density map, we distinguish the disk and tail regions as described in {\color{blue} {\bf Section \ref{sec:bound}}}.
The disk region within the $1\sigma$ boundary (red dashed ellipse) includes most of the Voronoi bins except for those in the eastern tail.
The SFR surface density is highly concentrated in the central region.
The flux-weighted mean of the Balmer decrement is 3.16, corresponding to a $V$-band extinction magnitude of 0.31 mag, which is the lowest value among our sample.
The total SFR is estimated to be 7.12 \Moyr, with 6.55 \Moyr~for the disk and 0.56 \Moyr~for the tail.
The tail SFR fraction of total SFR is $f_{\rm SFR}=7.9\%$, which is the lowest value out of the five jellyfish galaxies.

This galaxy shows a weak rotation of ionized gas in its disk. 
The line-of-sight velocity relative to the galactic center is $-63.7~{\rm km~s^{-1}}$ (towards the observer) in the northern star-forming knots and $45.1~{\rm km~s^{-1}}$ (away from the observer) in the eastern tail.
In the disk region, the maximum rotational velocity is estimated to be $54.4~{\rm km~s^{-1}}$ at a disk radius of $R_{\rm disk,c}\sim1\farcs4$.
The star-forming knots follow the disk rotation well, which is consistently shown in our sample.

The flux-weighted mean of the gas velocity dispersion is $\langle\sigma_{v,\rm{gas}}\rangle=30.0~{\rm km~s^{-1}}$, with $\langle\sigma_{v,\rm{gas}}\rangle=31~{\rm km~s^{-1}}$ for the disk and $\langle\sigma_{v,\rm{gas}}\rangle=15~{\rm km~s^{-1}}$ for the tail region.
In particular, the eastern tail region shows a very low gas velocity dispersion.
The low gas velocity dispersion of this galaxy implies that the star-forming regions are dynamically cold, which is consistent with the GASP studies.

\begin{figure*}
	\centering
	\includegraphics[width=\textwidth]{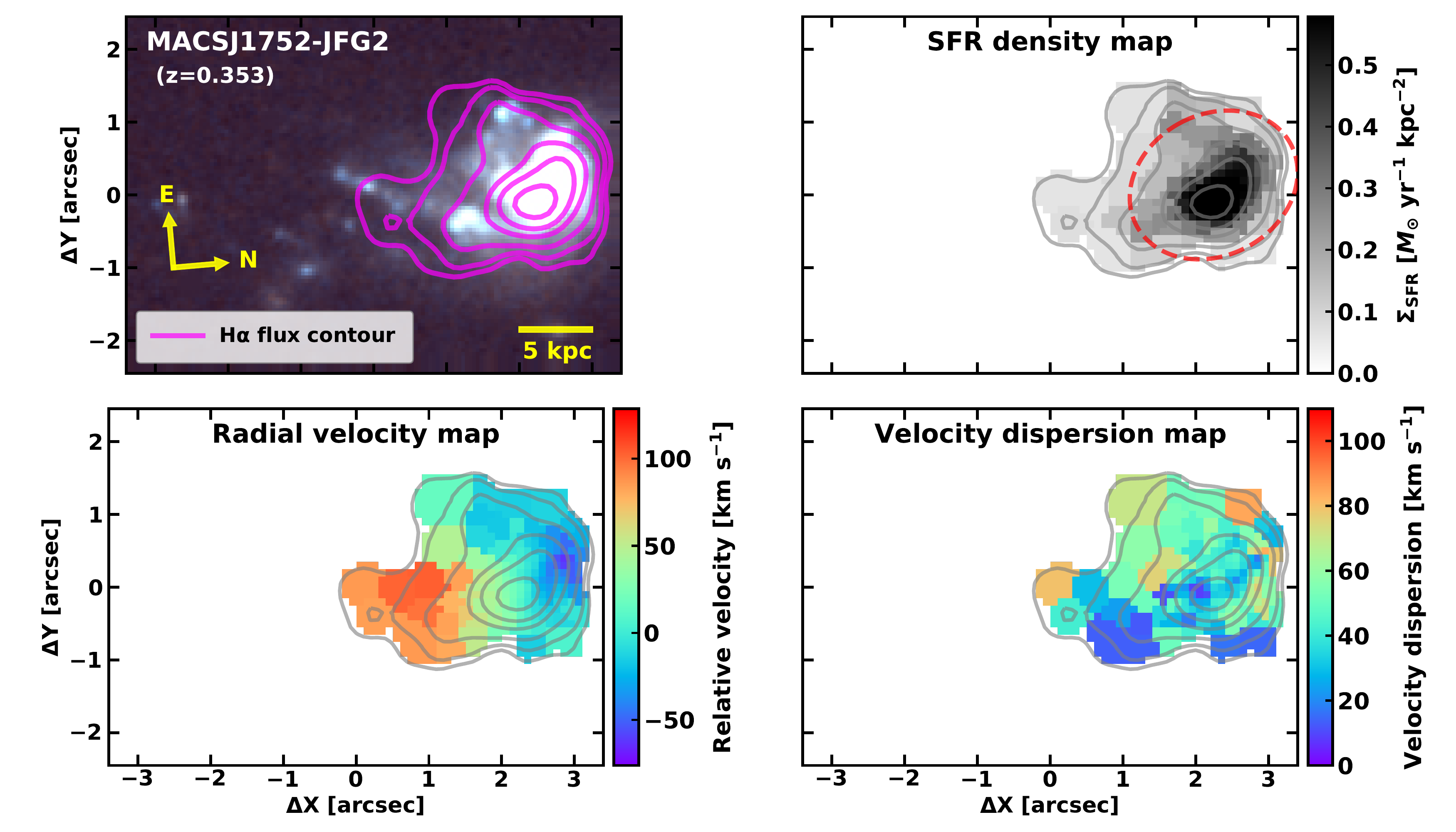}
	\caption{
	Same 
	as {\color{blue} {\bf Figure \ref{fig:IFU1}}} but for MACSJ1752-JFG2.
	\label{fig:IFU2}}
\end{figure*}

\begin{figure*}
	\centering
	\includegraphics[width=\textwidth]{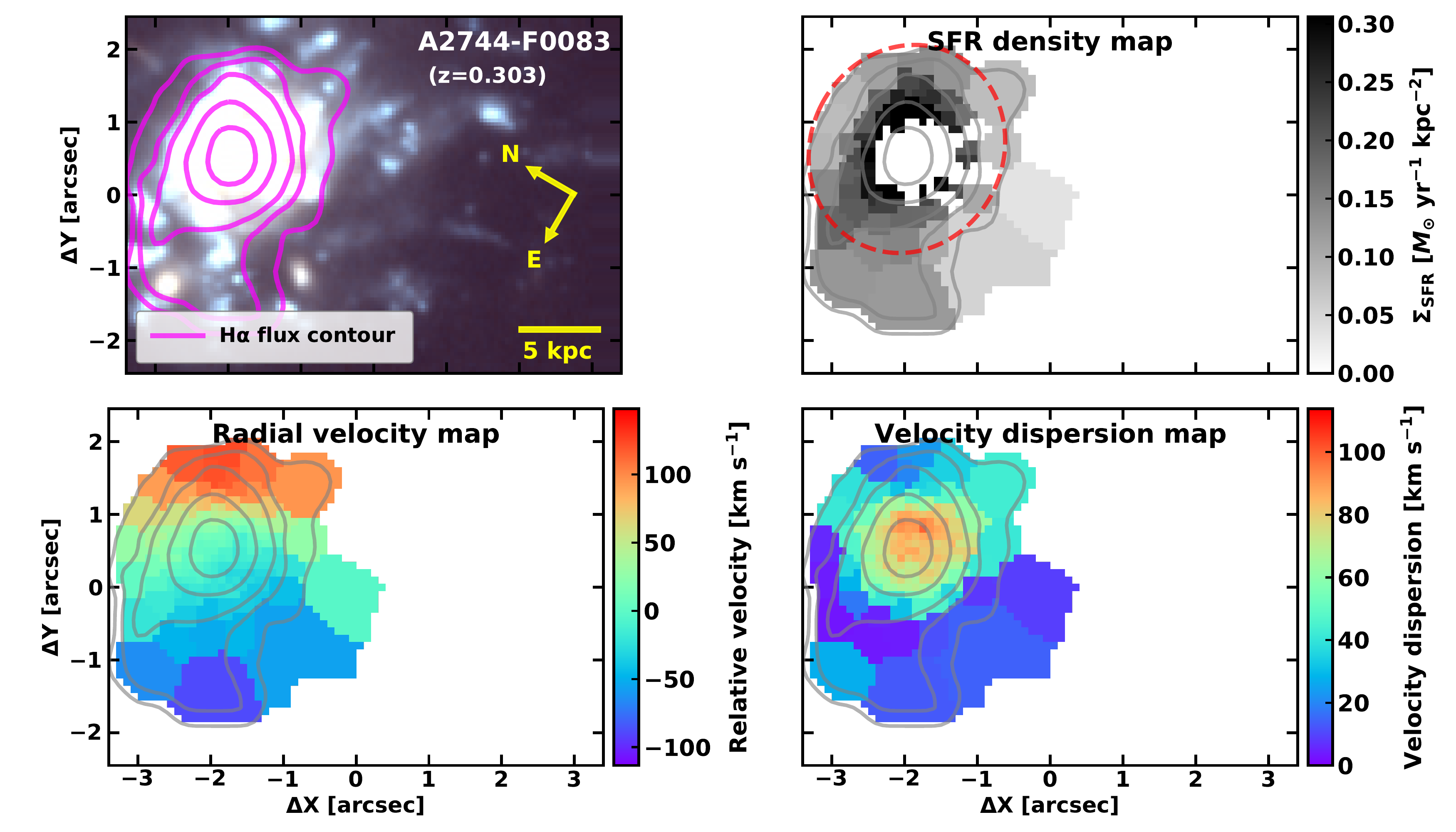}
	\caption{
	Same figure as {\color{blue} {\bf Figure \ref{fig:IFU1}}} but for A2744-F0083.
	\label{fig:IFU3}}
\end{figure*}

\begin{figure*}
	\centering
	\includegraphics[width=\textwidth]{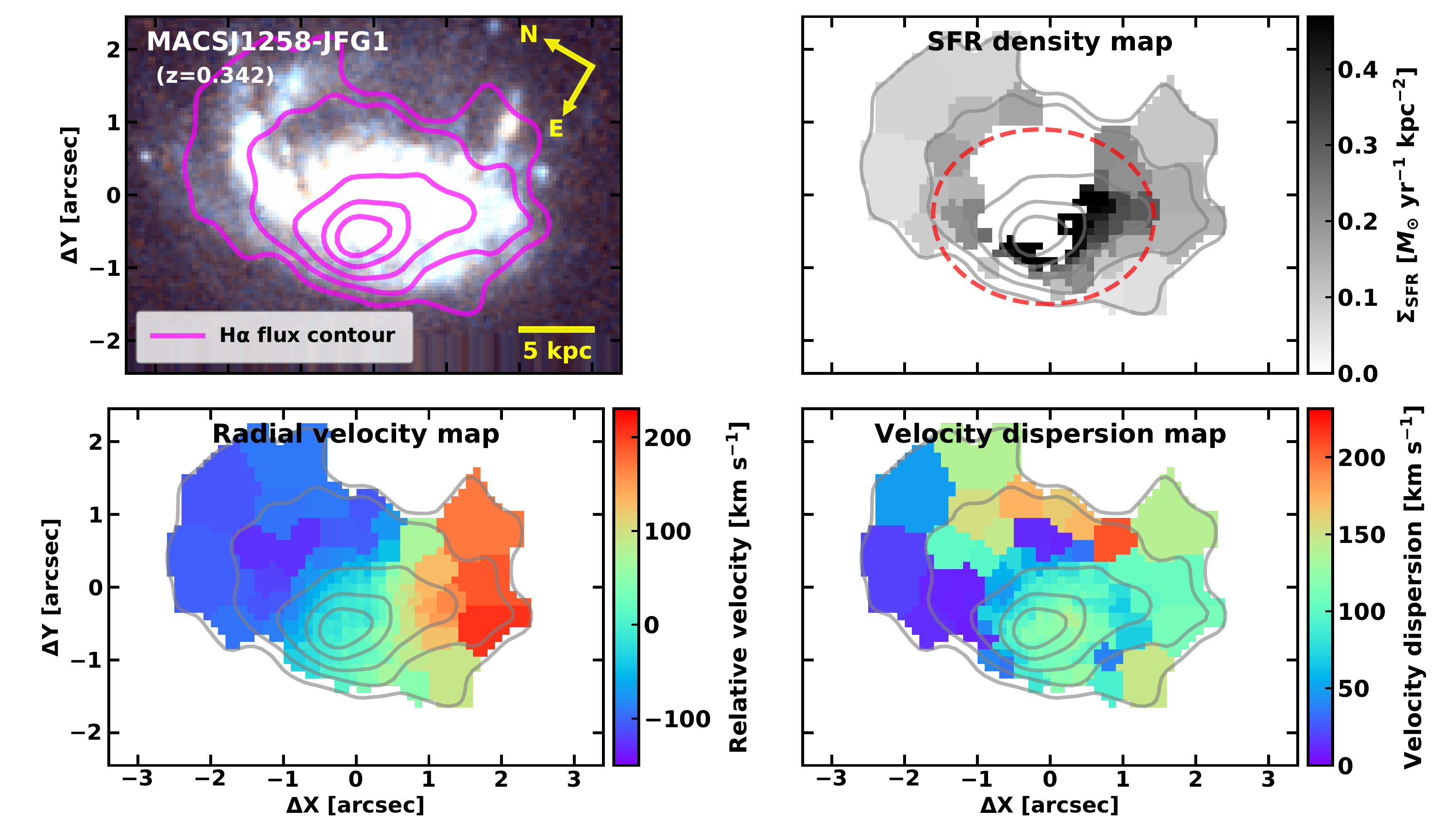}
	\caption{
	Same 
	as {\color{blue} {\bf Figure \ref{fig:IFU1}}} but for MACSJ1258-JFG1.
	\label{fig:IFU4}}
\end{figure*}

\begin{figure*}
	\centering
	\includegraphics[width=\textwidth]{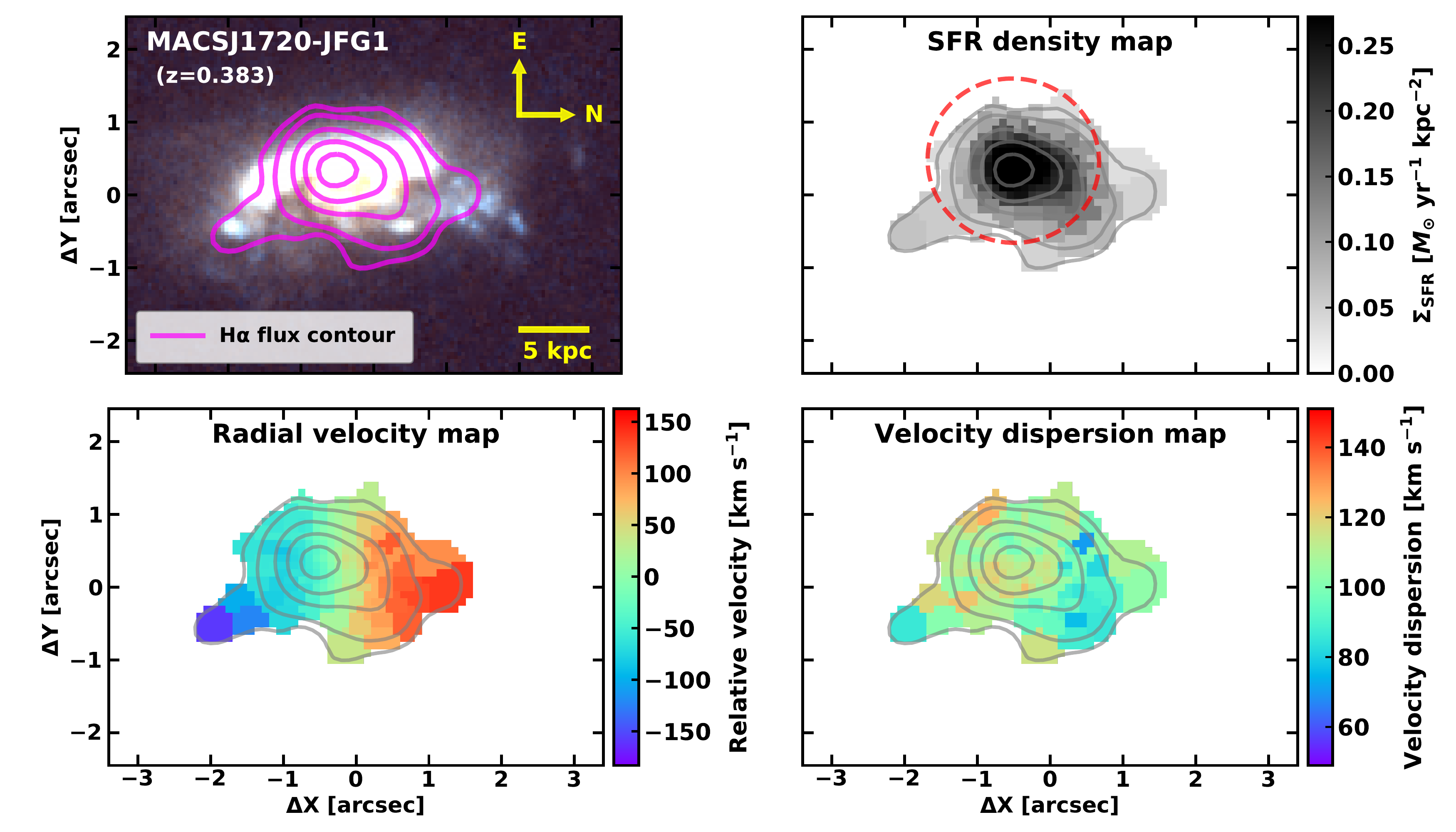}
	\caption{
	Same 
	as {\color{blue} {\bf Figure \ref{fig:IFU1}}} but for MACSJ1720-JFG1.
	\label{fig:IFU5}}
\end{figure*}

\subsubsection{MACSJ1752-JFG2}
In {\color{blue} {\bf Figure \ref{fig:IFU2}}}, MACSJ1752-JFG2 shows a long tail extending $\sim25~{\rm kpc}$ ($\sim5\arcsec$ in the angular scale) 
from the galactic center towards the southern direction.
In the \Ha~flux distribution, our GMOS/IFU data successfully covers the southern tail except for a few faint blue knots in the outer region.
This galaxy also shows two disturbed regions at a galactocentric distance of $\sim5~{\rm kpc}$ towards the south and east directions, exhibiting luminous blue knots.
The \Ha~flux distribution reproduces these peculiar substructures very well, which is indicative of ongoing star formation in the substructures. 

In the SFR surface density map, the ``tail'' region outside the $1\sigma$ boundary (red dashed ellipse) includes the southern tail and some star-forming knots in the east.
This galaxy shows the highest \Ha~to \Hb~ratio among our sample (${\rm H\alpha/H\beta=4.78}$), corresponding to a dust extinction magnitude of $1.61~{\rm mag}$.
Correcting for dust extinction, we obtain a total SFR of 30.43 \Moyr~(23.67 \Moyr~in the disk and 6.76 \Moyr~in the tail) with a tail SFR fraction of $f_{\rm SFR}=22.2\%$.
This indicates that the star formation activity is very strong in both the disk and the tail of this object.

MACSJ1752-JFG2 shows a disk rotation with a maximum rotational velocity of 73.2 \kms.
It seems that the substructures follow the disk rotation well, 
but the southern tail shows a high relative velocity of $\sim100$ \kms.
This can be a sign of ongoing RPS, indicating that the tail is stripped away from the center of the galaxy.

We obtain the flux-weighted mean of gas velocity dispersion of $\langle\sigma_{v,\rm{gas}}\rangle=43.1~{\rm km~s^{-1}}$, indicating that this galaxy is dynamically cold as well.
The mean velocity dispersions of the disk ($\langle\sigma_{v,\rm{gas}}\rangle=42~{\rm km~s^{-1}}$) and tail ($\langle\sigma_{v,\rm{gas}}\rangle=47~{\rm km~s^{-1}}$) do not show a significant difference.
However, there is some difference in the velocity dispersion between the two bright knots in the south and east $\sim5~{\rm kpc}$ from the center.
The southern large knot shows lower velocity dispersion ($\langle\sigma_{v,\rm{gas}}\rangle\sim25~{\rm km~s^{-1}}$) than the eastern one ($\langle\sigma_{v,\rm{gas}}\rangle\sim50~{\rm km~s^{-1}}$).
This implies that these two knots have different dynamical states, suggesting that ionized gas in the eastern knot is more turbulent than that in the southern one.

\subsubsection{A2744-F0083}
A2744-F0083 is the most spectacular galaxy, exhibiting numerous star-forming knots outside the main disk and non-ordered tails.
There are very bright star-forming knots at both the eastern and western sides of the disk.
These knots have comparable sizes and luminosities to compact dwarf galaxies \citep{owe12}.
The \Ha~flux distribution is highly asymmetric, peaking in the eastern region.
There are also long tails and extraplanar knots to $\sim25~{\rm kpc}$ ($\sim5\farcs5$) 
towards the southwestern direction from the center, but the \Ha~S/N of our GMOS/IFU data is very low in these regions.
This is because the observation program of A2744-F0083 used the R400 grating with higher spectral resolution and lower S/N for this galaxy.
Thus, we mainly analyze the ionized gas properties of the region within $\sim10~{\rm kpc}$ ($\sim2\farcs2$) from the center for this galaxy.

The SFR surface density map shows very strong star formation activity in the disk and in the eastern tail region.
Since we remove the AGN contribution to the SFR by subtracting the Voronoi bins with log([\ion{N}{2}]$\lambda6584/{\rm H\alpha})>-0.4$, several bins in the central region are excluded in the map.
For dust correction, we obtain a \Ha/\Hb~ratio of 4.14 from the AAOmega fiber spectra.
The total SFR is 22.00 \Moyr~with a disk SFR of 14.47 \Moyr~and a tail SFR of 7.53 \Moyr~($f_{\rm SFR}=34.2\%$).
The high SFR ratio in the tail indicates that the blue knots in the east undergo vigorous star formation.
For comparison, \citet{raw14} estimated the SFR of this galaxy to be 34.2 \Moyr~based on the sum of the UV and IR luminosities.
Different SFR indicators could result in this difference of SFRs because the \Ha~luminosity used in this study traces more recent star formation than the UV and IR luminosities \citep{ken12}.

The radial velocity map shows a clear disk rotation.
The maximum rotational velocity is 89.5 \kms~in the disk ($R_{\rm disk,c}\sim1\farcs4$).
Outside the disk, the star-forming knots in the eastern region show a radial velocity of $-88~{\rm km~s^{-1}}$.
The symmetric feature of the radial velocity distribution indicates that the star-forming regions follow the disk rotation well.

In the gas velocity dispersion map, we observe a clear radial gradient from the central disk ($\langle\sigma_{v,\rm{gas}}\rangle=54~{\rm km~s^{-1}}$) to the tail region ($\langle\sigma_{v,\rm{gas}}\rangle=18~{\rm km~s^{-1}}$).
This is because the \Ha~emission lines from the central disk are contaminated by the AGN activity, which broadens the line profiles.
This feature has been also shown in the GASP jellyfish galaxies hosting AGNs \citep{pog19}.
Star-forming knots in the tail region are dynamically cold, which is similar to those in MACSJ0916-JFG1 and MACSJ1752-JFG2.

\subsubsection{MACSJ1258-JFG1}
MACSJ1258-JFG1 is the most massive galaxy ($\log M_{\ast}/M_{\odot}=10.90$) among our sample and hosts a type I AGN.
In addition, it has a luminous tail as long as $\sim10~{\rm kpc}$ ($\sim2\farcs1$) in the northern region and plenty of blue knots are distributed around the disk.
There is also an asymmetric tail feature in the southwestern region of the galaxy.
Overall, these substructures seem to be consistent with the \Ha~flux distribution of this galaxy.

Like A2744-F0083, the SFR surface density map excludes the central region with the BPT classification of AGN or LINER.
We obtain a mean Balmer decrement of 4.64, which corresponds to a $V$-band extinction magnitude of 1.52 mag.
We measure a total SFR of 35.71 \Moyr, with a disk SFR of 18.91 \Moyr and a tail SFR of 16.80 \Moyr~($f_{\rm SFR}=47\%$).
The $f_{\rm SFR}$ value is the highest among our targets.
This high fraction of the tail SFR might result from strong star formation activity in the substructures in the northern and the southwestern regions ($\sim 10~{\rm kpc}$ from the center of the galaxy).

The radial velocity map shows a strong rotation with a maximum rotation velocity of 134.5 \kms~in the disk.
The northern tail shows a radial velocity of $\sim-100$ \kms~and the southern region shows a radial velocity of $\sim150$ \kms.
These substructures follow the disk rotation well.

The mean gas velocity dispersion of this galaxy is estimated to be $\langle\sigma_{v,\rm{gas}}\rangle=100.5~{\rm km~s^{-1}}$, which is higher than those of MACSJ0916-JFG1 and MACSJ1752-JFG2.
The mean gas velocity dispersion within the disk is $\langle\sigma_{v,\rm{gas}}\rangle=93.5~{\rm km~s^{-1}}$, which might be affected by AGN activity like A2744-F0083.
The northern tail shows a lower mean velocity dispersion with $\langle\sigma_{v,{\rm gas}}\rangle\sim50~{\rm km~s^{-1}}$, whereas the southern region shows a mean gas velocity dispersion of $\langle\sigma_{v,{\rm gas}}\rangle\sim90~{\rm km~s^{-1}}$.
This indicates that ionized gas in the northern tail is dynamically cold and that in the southern region is more turbulent than the northern region.

\subsubsection{MACSJ1720-JFG1}
MACSJ1720-JFG1 exhibits interesting substructures such as a bright arc-shaped region at the head of the disk and several blue extraplanar knots in the tail region.
The blue knots are located at both the north and south sides of the disk.
The \Ha~flux map from the GMOS/IFU data reproduces these optical features well.

The SFR surface density map shows that the distribution of star formation activity is strongly concentrated at the the center of the galaxy.
We estimated a mean Balmer decrement, obtaining \Ha/\Hb$=4.05\pm1.40$ ($A_V=1.09~{\rm mag}$).
We derive a total SFR of 17.37 \Moyr, with a disk SFR of 14.09 \Moyr and a tail SFR of 3.28 \Moyr~($f_{\rm SFR}=19\%$).

The radial velocity map shows a clear disk rotation on the axis of the east-west direction.
The maximum rotation velocity within the disk is 122.0 \kms.
The northern tail shows a radial velocity of $\sim120$ \kms~and the blue blob in the southern region shows a radial velocity of $\sim-100$ \kms.
These disturbed features also rotate following the disk of this galaxy, which is similar to other jellyfish galaxies.

This galaxy shows a mean gas velocity dispersion of $\langle\sigma_{v,\rm{gas}}\rangle=104.5~{\rm km~s^{-1}}$, which is much higher than those of MACSJ0916-JFG1 and MACSJ1752-JFG2, but similar to that of MACSJ1258-JFG1.
This indicates that ionized gas in the star-forming knots of this galaxy seems to be more turbulent than that in the star-forming regions in other jellyfish galaxies.
Other shock-heating mechanisms are likely to contribute to the gas ionization in this galaxy, as also suggested by its location in the composite region in the BPT diagram (see {\color{blue} {\bf Figure \ref{fig:BPT1}}}). 


\begin{figure*}
	\centering
	\includegraphics[width=\textwidth]{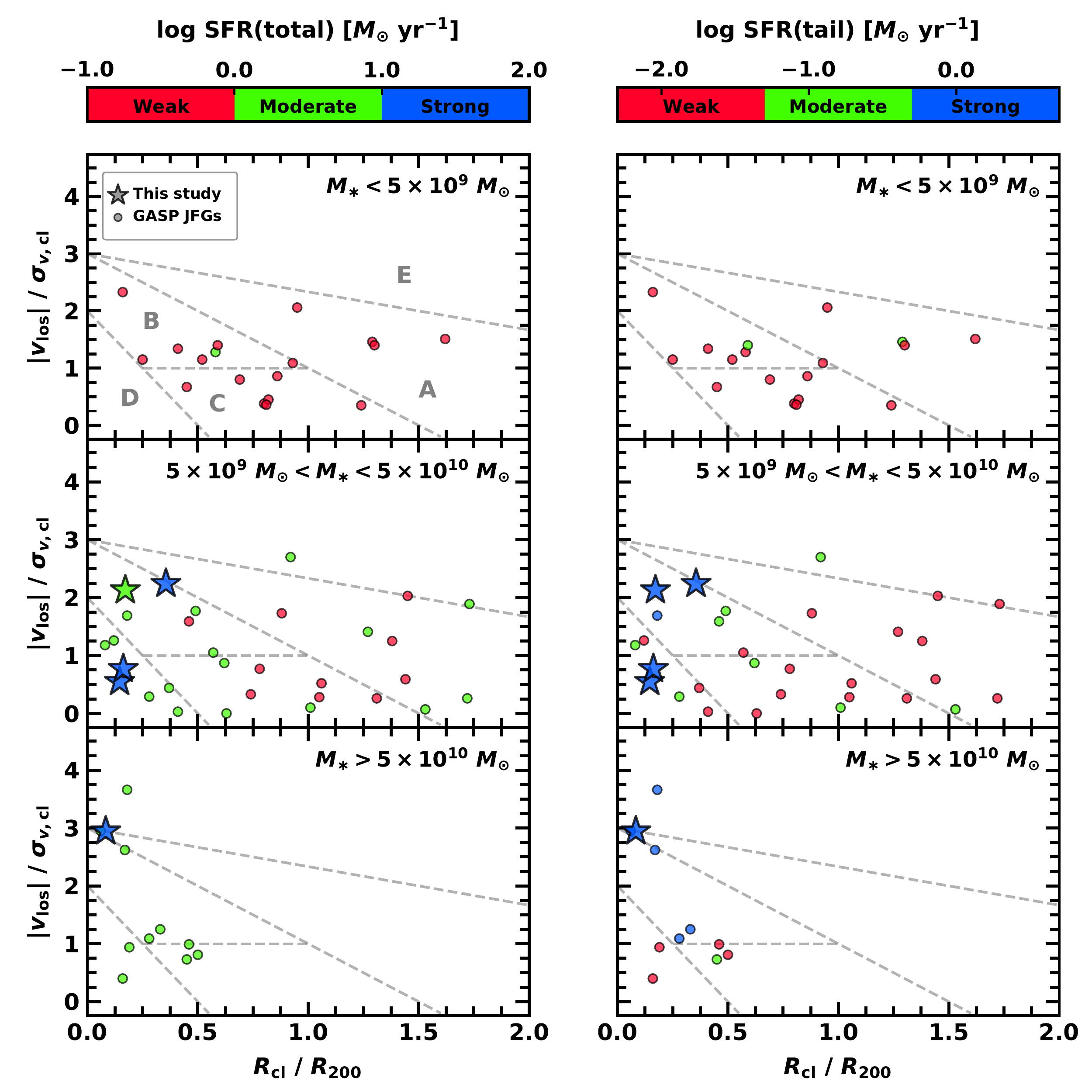}
	\caption{
	Projected phase-space diagrams of our sample (star symbols) and the GASP jellyfish galaxies (circles).
	We normalize clustercentric distance ($R_{\rm cl}$) and absolute relative velocity ($|v_{\rm los}|$) with cluster virial radius ($R_{200}$) and velocity dispersion ($\sigma_{v, {\rm cl}}$), respectively.
	All the data are color-coded by total SFR (left) and tail SFR (right).
	The color bars on the top denote the logarithmic scale of each SFR.
	Gray dashed lines represent the boundaries of the five regions that were roughly defined by infall stages of cluster galaxies \citep{rhe17, mun21}: Region A (first infall), B (recent infall), C (intemediate infall), D (ancient infall), and E (field).
	We divide the whole sample into three categories by stellar mass: low-mass ($M_{\ast}<5\times10^{9}~M_{\odot}$; upper), intermediate-mass ($5\times10^{9}~M_{\odot}<M_{\ast}<5\times10^{10}~M_{\odot}$; middle), and high-mass ($M_{\ast}>5\times10^{10}~M_{\odot}$; lower) galaxies.
	\label{fig:PSD1}}
\end{figure*}

\begin{figure*}
	\centering
	\includegraphics[width=\textwidth]{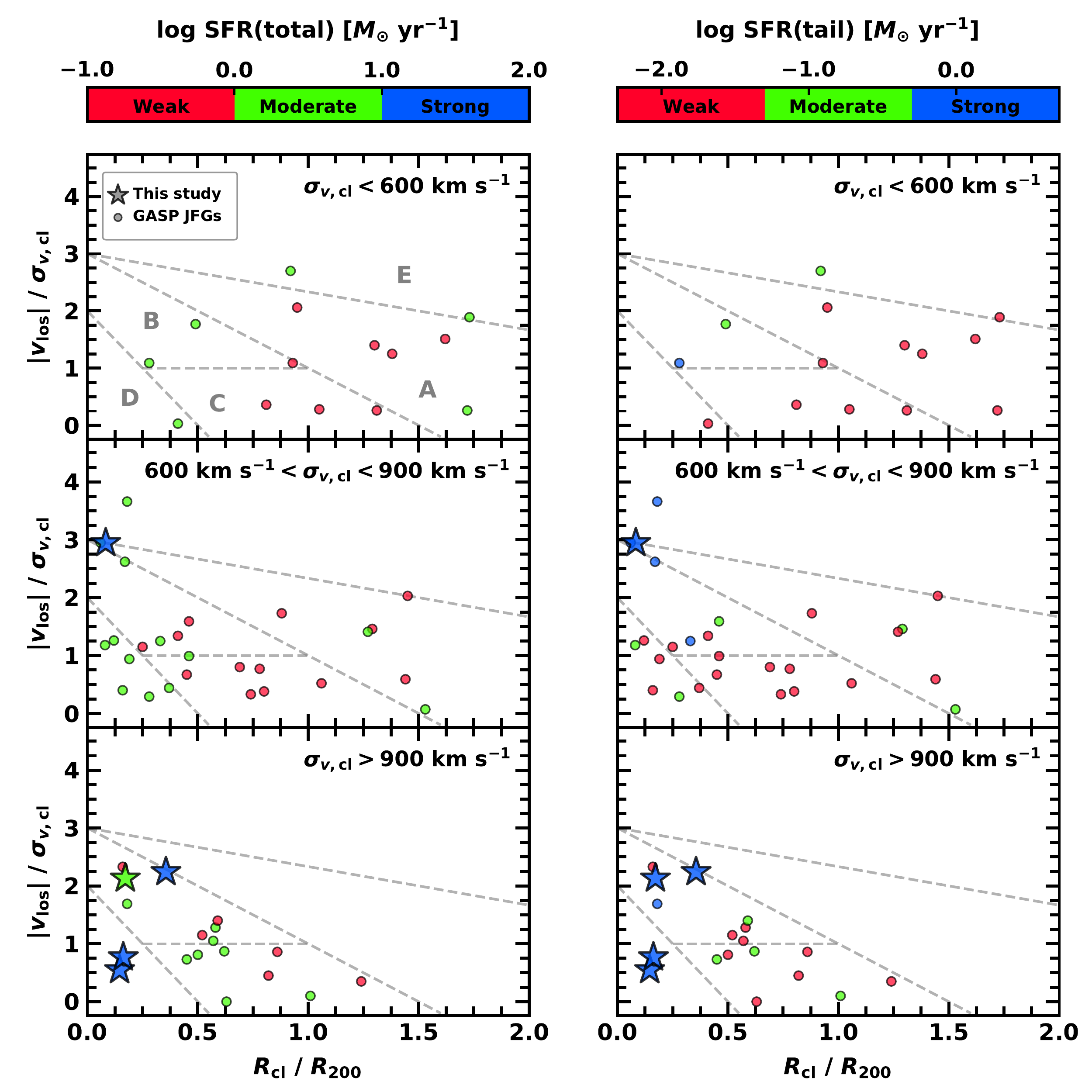}
	\caption{
	Same 
	as {\color{blue} {\bf Figure \ref{fig:PSD1}}}, but we divide the sample into three categories by cluster velocity dispersion: low-mass ($\sigma_{v, {\rm cl}}<600~{\rm km~s^{-1}}$; upper), intermediate-mass ($600~{\rm km~s^{-1}}<\sigma_{v, {\rm cl}}<900~{\rm km~s^{-1}}$; middle), and high-mass ($\sigma_{v, {\rm cl}}>900~{\rm km~s^{-1}}$; lower) clusters.
	\label{fig:PSD2}}
\end{figure*}

\section{Phase-space Diagrams of the Jellyfish Galaxies}
\label{sec:psd}

Phase-space diagrams are useful to understand the impact of environmental effects on cluster galaxies concerning their orbital histories \citep{rhe17, jaf18, mun21}.
We display the projected phase-space diagrams of our targets and the GASP jellyfish galaxies \citep{gul20} by categorizing the sample based on stellar mass ({\color{blue} {\bf Figure \ref{fig:PSD1}}}) and host cluster velocity dispersion ({\color{blue} {\bf Figure \ref{fig:PSD2}}}).
\citet{lee22} showed the phase-space diagrams of the jellyfish galaxies with different categories of jellyfish morphology.
We plot the 2D clustercentric distance normalized by the virial radius of the host cluster ($R_{\rm cl}/R_{200}$) on the x-axis, and we plot the velocity relative to the cluster normalized by the cluster velocity dispersion ($|\Delta v_{\rm los}|/\sigma_{v,{\rm cl}}$) on the y-axis.
We measure the line-of-sight velocity of each galaxy with $\Delta v_{\rm los}=c\times(z_{\rm gal}-z_{\rm clu})/(1+z_{\rm clu})$ where $c$ is the speed of light, $z_{\rm gal}$ is the galaxy redshift, and $z_{\rm clu}$ is the cluster redshift.

In {\color{blue} {\bf Figures \ref{fig:PSD1}}} and {\color{blue} {\bf \ref{fig:PSD2}}}, we display total SFRs (left panels) and tail SFRs (right panels) at the top of each panel by categorizing the star formation activity:  
``weak'' (red), ``moderate'' (green), and ``strong'' (blue) star formation.
Gray dashed lines denote the regions (from A to E) classified in \citet{rhe17} and \citet{mun21}, showing the approximate stages of galaxy infall to the clusters: Region A (first infall), B (recent infall), C (intermediate infall), D (ancient infall), and E (field).
In this study, we classify the location of jellyfish galaxies as ``inner region'' ($R_{\rm cl}/R_{200}\leq0.5$) and ``outer region'' ($R_{\rm cl}/R_{200}>0.5$) in their host clusters.

In {\color{blue} {\bf Figure \ref{fig:PSD1}}}, we plot phase-space diagrams for different bins of galaxy stellar mass: low-mass ($M_{\ast}<5\times10^{9}~M_{\odot}$), intermediate-mass ($5\times10^{9}~M_{\odot}<M_{\ast}<5\times10^{10}~M_{\odot}$), and high-mass ($M_{\ast}>5\times10^{10}~M_{\odot}$).
In the low-mass regime, almost all GASP jellyfish galaxies show weak star formation activity (17/18 in total SFR and 16/18 in tail SFR).
They are primarily located in the outer region ($R_{\rm cl}/R_{200}>0.5$), with only four of them in the inner region ($R_{\rm cl}/R_{200}\leq0.5$).
Note that no galaxies are located in the ancient infall region, implying that the low-mass jellyfish galaxies might be in the early stages of cluster infall.

In the intermediate-mass regime, the GASP jellyfish galaxies show a wide range of clustercentric distances and velocities in their host clusters.
These galaxies are also primarily located in the outer region.
Most of them show weak star formation activity in terms of tail SFRs.
In contrast, four jellyfish galaxies in this study are located in the inner region of the clusters, and they show moderate or strong star activity.
With the combined sample of the GASP survey and this study, this panel shows that the fraction of galaxies with weak star formation activity is higher in the outer region (9/18 in total SFR and 15/18 in tail SFR) than in the inner region (1/12 in total SFR and 4/12 in tail SFR).

In the high-mass regime, there are 10 GASP jellyfish galaxies and MACSJ1258-JFG1 in this study.
All massive jellyfish galaxies are located in the inner region of the host clusters.
Most of the jellyfish galaxies show moderate or strong star formation activity.
The panels in this figure indicates that jellyfish galaxies with higher stellar mass and lower clustercentric distances are likely to exhibit stronger star formation activity on both global (total SFR) and local (tail SFR) scales.
This trend has also been observed in \citet{gul20} (see their Figure 4).


In {\color{blue} {\bf Figure \ref{fig:PSD2}}}, we plot the phase-space diagrams in different bins of cluster velocity dispersion: low-mass hosts ($\sigma_{v,{\rm cl}}<600~{\rm km~s^{-1}}$), intermediate-mass hosts ($600~{\rm km~s^{-1}}<\sigma_{v,{\rm cl}}<900~{\rm km~s^{-1}}$), and high-mass hosts ($\sigma_{v,{\rm cl}}>900~{\rm km~s^{-1}}$).
For the low-mass host clusters, half of the 14 GASP galaxies are located outside the virial radius.
A majority of the jellyfish galaxies show weak star formation activity, but galaxies in the inner region mostly show stronger star formation activity than those in the outer region.
This trend can be also seen in intermediate-mass host clusters.
Most galaxies in the outer region show weak star formation activity (9/11 in total and tail SFR), whereas those in the inner region have a lower fraction of weak star formation activity (4/16 in total SFR and 9/16 in tail SFR).
For the high-mass host clusters, there are 14 GASP jellyfish galaxies and 4 galaxies in this study.
In terms of total SFRs, the fraction of galaxies with moderate and strong star formation activity (12/18) is higher in high-mass clusters than in low-mass (6/14) and intermediate-mass (13/27) clusters.
This trend is also seen with tail SFRs: 3/14 in low-mass, 9/27 in intermediate-mass, and 8/18 in high-mass clusters.
This indicates that the star formation activity of jellyfish galaxies in high-mass clusters is likely to be more enhanced compared to those in low-mass and intermediate-mass clusters.
In addition, there are no GASP galaxies located in the first infall region (``A'') in these massive clusters in contrast to low- and intermediate-mass clusters.
This implies that the jellyfish galaxies in massive clusters are at a later phase of cluster infall than in low-mass clusters.

Combining the GASP sample and our sample, these phase-space diagrams reveal the following.
First, our jellyfish galaxy sample is located in the inner region with a wide range of relative velocities.
Second, it is clearly shown that jellyfish galaxies in the inner region tend to have higher SFRs than those in the outer region.
Third, the star formation activity of jellyfish galaxies tends to increase as galaxy stellar mass increases, as shown in {\color{blue} {\bf Figure \ref{fig:PSD1}}}.
This reflects the SFR-mass relation of jellyfish galaxies as also shown in Figure 2 and 4 in \citet{lee22}.

Finally, the star formation activity of jellyfish galaxies tend to increase as the host cluster velocity dispersion increases, as shown in {\color{blue} {\bf Figure \ref{fig:PSD2}}}.
In this context, \citet{lee22} revealed that jellyfish galaxies with strong RPS signatures show more enhanced star formation activity in more massive clusters with a higher degree of RPS.

\section{Summary}
\label{sec:summary}

In this study, we 
presented an observational study of the five jellyfish galaxies in the MACS clusters and Abell 2744 at $z>0.3$ 
based on Gemini GMOS/IFU observations.
Using these data, we investigated the ionized gas properties of these jellyfish galaxies such as ionization mechanisms, kinematics, and SFRs.
\citet{lee22} also used the \Ha-derived SFRs in this study and suggested positive correlations between the star formation activity of jellyfish galaxies and their host cluster properties.
Our main results can be summarized as follows.

\begin{enumerate}
	\item The BPT diagrams of [\ion{O}{3}]$\lambda$5007/\Hb~and [\ion{N}{2}]$\lambda$6584/\Ha~show that the five jellyfish galaxies have different ionization mechanisms.
    MACSJ0916-JFG1 and MACSJ1752-JFG2 are located in the star-forming region, indicating that the gas contents in these galaxies are ionized purely by star formation.
	On the other hand, A2744-F0083 and MACSJ1258-JFG1 are located in the AGN region with high line ratios of [\ion{O}{3}]$\lambda$5007/\Hb~and [\ion{N}{2}]$\lambda$6584/\Ha.
	MACSJ1720-JFG1 is located in the composite region, implying a mixed contribution of star formation and other shock-heating mechanisms.
	\item The spatial distributions of the \Ha~flux are well-matched with the optical features in all five jellyfish galaxies.
	This indicates that the ionized gas distribution is consistent with that of the stellar light distribution in the jellyfish galaxies.
	\item The radial velocity distributions of the jellyfish galaxies indicate that ionized gas in the disk and tail regions rotates around the center of each galaxy.
	Some tail regions (e.g.~eastern side of MACSJ1752-JFG2) show high relative velocities with respect to the center of the galaxy, which indicates signs of RPS.
	\item MACSJ0916-JFG1, MACSJ1752-JFG2 and the tail regions in A2744-F0083 and MACSJ1258-JFG1 show a mean velocity dispersion lower than $50~{\rm km~s^{-1}}$, which is consistent with the mean value of star-forming clumps in the GASP jellyfish galaxies.
	This implies that the ionized gas in those regions is dynamically cold.
	In contrast, MACSJ1720-JFG1 and the central regions in A2744-F0083 and MACSJ1258-JFG1 show mean velocity dispersions higher than $50~{\rm km~s^{-1}}$, indicating that the ionized gas is more turbulent than typical star-forming regions.
	This could be associated with the AGN activity or other shock-heating mechanisms.
	\item The total and tail SFRs of the five jellyfish galaxies are much higher than those of the GASP sample.
	The median SFRs of our targets are 22.0 \Moyr~in total and 6.8 \Moyr~in the tails, whereas those of the GASP sample are 1.1 \Moyr~in total and 0.03 \Moyr~in the tails.
	In addition, the median SFR fraction in the tail ($f_{\rm SFR}$) is also much higher in this study ($f_{\rm SFR}=22\%$) than in the GASP studies ($f_{\rm SFR}=3\%$).
	\item In the projected phase-space diagrams, the jellyfish galaxies in this study are located in the inner region with a wide range of orbital velocities relative to the cluster center.
	Combining the GASP sample and our sample, we find that jellyfish galaxies with higher stellar masses and higher host cluster velocity dispersions are more likely to be located in the inner region of the clusters with more enhanced star formation activity.
\end{enumerate}

We thank the anonymous referee for his/her important comments and suggestions to improve the submitted manuscript.
This study was supported by the National Research Foundation grant funded by the Korean Government (NRF-2019R1A2C2084019).
This work was supported by K-GMT Science Program (PID: GS-2019A-Q-214, GN-2019A-Q-215, GS-2019B-Q-219, and GN-2021A-Q-205) of Korea Astronomy and Space Science Institute (KASI).
Based on observations obtained at the international Gemini Observatory, a program of NSF’s NOIRLab, acquired through the Gemini Observatory Archive at NSF’s NOIRLab and processed using the Gemini IRAF package, which is managed by the Association of Universities for Research in Astronomy (AURA) under a cooperative agreement with the National Science Foundation on behalf of the Gemini Observatory partnership: the National Science Foundation (United States), National Research Council (Canada), Agencia Nacional de Investigaci\'{o}n y Desarrollo (Chile), Ministerio de Ciencia, Tecnolog\'{i}a e Innovaci\'{o}n (Argentina), Minist\'{e}rio da Ci\^{e}ncia, Tecnologia, Inova\c{c}\~{o}es e Comunica\c{c}\~{o}es (Brazil), and Korea Astronomy and Space Science Institute (Republic of Korea).
This research has made use of the NASA/IPAC Extragalactic Database (NED),
which is operated by the Jet Propulsion Laboratory, California Institute of Technology,
under contract with the National Aeronautics and Space Administration. 
This research is based on observations made with the NASA/ESA Hubble Space Telescope and the Galaxy Evolution Explorer obtained from the Mikulski Archive for Space Telescopes (MAST) data archive at the Space Telescope Science Institute (STScI).
STScI is operated by the Association of Universities for Research in Astronomy, Inc., under NASA contract NAS 5-26555.
This publication makes use of data products from the Wide-field Infrared Survey Explorer, which is a joint project of the University of California, Los Angeles, and the Jet Propulsion Laboratory/California Institute of Technology, funded by the National Aeronautics and Space Administration.
This work is based in part on observations made with the Spitzer Space Telescope, which was operated by the Jet Propulsion Laboratory, California Institute of Technology under a contract with NASA.

\software{Numpy \citep{har20}, Matplotlib \citep{hun07}, Scipy \citep{vir20}, Astropy \citep{ast13, ast18}, PyRAF \citep{sts12}, Vorbin \citep{cap03}, Emcee \citep{for13}, Photutils \citep{bra21}, and SExtractor \citep{ber96}}

\clearpage

\end{document}